\documentclass[11pt]{article}

\usepackage{amsfonts,amsmath,amssymb,mathrsfs,amsthm}
\usepackage{epsfig,epstopdf,graphicx,graphics}
\usepackage{color}
\usepackage{float}
\usepackage{ifpdf}
\usepackage[colorlinks=true]{hyperref}
\usepackage{braket}

\newtheorem{theorem}{Theorem}

\newtheorem{corollary}{Corollary}

\newtheorem{definition}{Definition}
\newtheorem{example}{Example}

\newtheorem{lemma}{Lemma}

\newtheorem{problem}{Problem}
\newtheorem{proposition}{Proposition}
\newtheorem{remark}{Remark}

\newcommand{\bthm}{\begin{theorem}}
\newcommand{\ethm}{\end{theorem}}
\newcommand{\blem}{\begin{lemma}}
\newcommand{\elem}{\end{lemma}}
\newcommand{\bex}{\begin{example}}
\newcommand{\eex}{\end{example}}
\newcommand{\bprop}{\begin{proposition}}
\newcommand{\eprop}{\end{proposition}}
\newcommand{\bplm}{\begin{problem}}
\newcommand{\eplm}{\end{problem}}
\newcommand{\bmrk}{\begin{remark}}
\newcommand{\emrk}{\end{remark}}
\newcommand{\bdfn}{\begin{definition}}
\newcommand{\edfn}{\end{definition}}
\newcommand{\bcor}{\begin{corollary}}
\newcommand{\ecor}{\end{corollary}}

\newcommand{\beq}{\begin{equation}}
\newcommand{\eeq}{\end{equation}}
\newcommand{\beqm}{\begin{equation*}}
\newcommand{\eeqm}{\end{equation*}}
\newcommand{\beqn}{\begin{eqnarray}}
\newcommand{\eeqn}{\end{eqnarray}}
\newcommand{\beqnm}{\begin{eqnarray*}}
\newcommand{\eeqnm}{\end{eqnarray*}}
\newcommand{\bea}{\begin{align}}
\newcommand{\eea}{\end{align}}
\newcommand{\beam}{\begin{align*}}
\newcommand{\eeam}{\end{align*}}

\newcommand{\bei}{\begin{itemize}}
\newcommand{\eei}{\end{itemize}}
\newcommand{\bed}{\begin{description}}
\newcommand{\eed}{\end{description}}
\newcommand{\bee}{\begin{enumerate}}
\newcommand{\eee}{\end{enumerate}}

\newcommand{\bey}{\begin{array}}
\newcommand{\eey}{\end{array}}

\newcommand{\beb}{\begin{thebibliography}{99}}
\newcommand{\eeb}{\end{thebibliography}}

\newcommand{\la}{\label}

\begin{document}

\title{On the dynamics of a quantum coherent feedback network of cavity-mediated double quantum dot qubits}

\author{Zhiyuan~Dong \thanks{School of Mechanical Engineering and Automation, Harbin Institute of Technology (Shenzhen), Shenzhen, China (dongzhiyuan@hit.edu.cn).} \and Wei~Cui\thanks{School of Automation Science and Engineering, South China University of Technology, Guangzhou, China (aucuiwei@scut.edu.cn).}, \and Guofeng~Zhang\thanks{
Guofeng Zhang is with the Department of Applied Mathematics, The Hong Kong Polytechnic University, Hong Kong  (e-mail: guofeng.zhang@polyu.edu.hk).}}
\maketitle  

\begin{abstract}
The purpose of this paper is to present a comprehensive study of a coherent feedback network  where the main component consists of two distant double quantum dot (DQD) qubits which are directly coupled to a cavity. This main component has recently been physically realized (van Woerkom, {\it et al.}, Microwave photon-mediated interactions between semiconductor qubits, Physical Review X, 8(4):041018, 2018).  The feedback loop is closed by cascading this main component with a beamsplitter. The dynamics of this coherent feedback network is studied from three perspectives. First, an analytic form of the output single-photon state of the network driven by a single-photon state is derived; in particular, it is observed that coherent feedback  elongates considerably the interaction between the input single photon and the network. Second, excitation probabilities of DQD qubits are computed when the network is driven by a single-photon input state. Moreover, if the input is vacuum but one of the two DQD qubits is initialized in its excited state, the explicit expression of the state of the network is derived, in particular, it is shown that the output field and the two DQD qubits can form an entangled state if the transition frequencies of two DQD qubits  are equal. Finally, the exact form of the pulse shape is obtained by which the single-photon input can fully excite one of these two  DQD qubits at any controllable time, which may be useful in the construction of $2$-qubit quantum gates.
 
\textbf{keywords.} single-photon state, double quantum dot qubit, quantum coherent feedback network, quantum control, open quantum systems
\end{abstract}

\section{Introduction}\label{sec:intro}

In the past few decades, quantum control has attracted much attention due to the rapid development of quantum information science and technology. Efficient manipulation of the interaction between photons (flying qubits) and finite-level quantum systems (stationary qubits) is necessary for quantum control which enables quantum communication \cite{GT07}, quantum network \cite{K08} and quantum filtering \cite{BvHJ07}. With the advancement of technology in quantum optics, the ultrastrong coupling regime of quantum light-matter interaction is currently an active research field \cite{braumuller2017,Lv18,FLRKS19}. The interaction between a DQD qubit and a nearby quantum point contact (QPC) is investigated in \cite{Cong15,CD18}. Particularly, Lyapunov-based control method is used in \cite{Cong15} to transfer the charge qubit to its target state. In \cite{CD18}, the master equation for a DQD qubit is derived and the measurement-induced backaction is considered. Moreover, a Hamiltonian feedback control law is proposed to realize and stabilize the current convergent to the target value.

As the flying qubits, single photons are a promising candidate for quantum information processing. For example, the strong nonlinear interaction between photons and optical emitters can be used to engineer a single-photon transistor \cite{CSDL07}. The operation principle of the single-photon transistor is to use either zero or one photon in the storage step, then the subsequent transmitted or reflected photons are controlled by the conditional flip of the ``gate" pulse. Another single-photon transistor is introduced in \cite{NLH13} to setup a circuit quantum electrodynamical (circuit QED) model, which consists of two two-level systems. Although no photons are exchanged between the two transmission lines in this circuit, one photon can completely block or enable the propagation of the other by the interaction between the two two-level systems. Recently, the realization of an optical transistor is given in \cite{CBB13}, which consists of a four-level system and a stored photon to control the transmission of source photons.

From a control-theoretic point of view, analysis of quantum systems' response to single-photon states is an essential aspect of control systems engineering. The interaction of quantum systems with single-photon states has been extensively studied, see e.g., \cite{SF05,FKS10,RF12,ZJ13,PZJ16,Z20}. The transmission and reflection probabilities in terms of the stationary output photon state are discussed by using the scattering matrix \cite{SF05,FKS10,RF12}. In \cite{PZJ16}, an analytical expression of the output field state is derived for a class of quantum finite-level systems driven by single-photon input states. Interestingly, it is shown that quantum linear systems theory \cite{ZJ13,Zhang14} can be adopted to derive the pulse shapes of the output single-photon states. On the other hand, the problem of quantum filtering for systems driven by single-photon states has been attracting growing interest due to their promising applications in quantum communication and measurement feedback control, see e.g., \cite{GJNC12,CHJ12,ZJ12,ZLWJN17,DZA18} and references therein.

Recently, the microwave photon-mediated interactions between semiconductor qubits have been physically implemented in \cite{Van18,BPB19,AES19}.  In this paper, we study an open quantum system which has recently been physically implemented on a semiconducting platform  \cite{Van18}.  In this system, two double quantum dots (DQDs) are modeled as two charge qubits. These two DQD qubits are separated from one another; however they are both directly coupled to a microwave cavity.  In other words, the cavity enables information exchange between the {\em distant} two DQD qubits, which is important for quantum information processing.  In \cite{Van18}, the dynamics of this system is studied when it is driven by a laser (namely a coherent state). In this paper, we are interested in its dynamics when it is driven by a single photon. For easy reference,  this system is called the coupled system $G$ in our paper. Moreover, we cascade a beamsplitter with $G$ to form a quantum coherent feedback network as shown in Fig. \ref{system_Jan16}, and aim to study the dynamics of this coherent feedback network. The contribution of this paper is three-fold as summarized below.

Firstly, an analytic expression of the output single-photon state is derived when the coherent feedback network is driven by a continuous-mode single-photon state; see Theorem \ref{theorem3.1}. To establish this result, the Routh like table and the Sign Pair Criterion (SPC) developed in \cite{SS12} are utilized. Moreover, techniques for single photon processing developed in \cite{ZJ13,PZJ16} are also employed. Theorem \ref{theorem3.1} is illustrated by using a single photon of an exponentially rising pulse shape to drive the coherent feedback network.  Differences among red detuned, blue detuned, and red+blue detuned dynamics are demonstrated. In particular, it is observed that coherent feedback is able to elongate considerably the interaction between the input single photon and the system.

Secondly, the excitation of the DQD qubits by a single-photon input state is investigated. In particular, it is demonstrated that red+blue detunings allow higher excitation probabilities. Moreover, assuming that one DQD qubit  is initialized in its excited state while the coherent feedback network is driven by a vacuum input, an explicit form of the state of the coherent  feedback network is derived by means of the quantum stochastic Schr\"{o}dinger equation (QSSE). In particular, when the transition frequencies of the two DQD qubits are equal to each other, it is shown that the output field and the two DQD qubits form an entangled state; see Theorem \ref{alpha}. This interesting phenomenon cannot occur if there is only one DQD qubit in this coherent feedback network.

Finally,  we study the problem of how to fully excite a DQD qubit by using a single photon with a special designed pulse shape; see Theorem \ref{thm:full inversion}.  To derive this result, both the Schr\"{o}dinger picture and the Heisenberg picture of open quantum systems have to be used together. A related problem is studied in \cite{PZCJ15}, where it is shown how to design a single-photon pulse shape to excite an atom residing in a cavity (Cavity QED).

The rest of this paper is organized as follows. Some preliminaries are summarized in Section \ref{sec:pre}, which include notation to be used, open quantum systems and single-photon states. The quantum coherent feedback network is presented in Section \ref{sec:model}. The steady-state output field state of the coherent feedback network driven by a single-photon input is derived in Section \ref{sec:state}. Section \ref{sec:excitation} presents the master equations for the 1st DQD qubit and discusses the changes of excitation probability with various system parameters. A single-photon inverting pulse is designed in Section \ref{sec:inverse}, which is able to fully excite a DQD qubit. Section \ref{sec:conclusion} concludes this paper.

\section{Preliminaries}\label{sec:pre}

In this section, we introduce the notation to be used in this paper. A concise introduction to open Markovian quantum systems and continuous-mode single-photon states is also provided, as the model studied in this paper lies in this framework.

\emph{Notation}. Let $i=\sqrt{-1}$ be the imaginary unit and $\ket{\Phi_0}$ the vacuum state of a free-propagating field. Given a column vector of complex numbers or operators $X=[x_1,\cdots,x_n]^\top$, the complex conjugate  or  adjoint operator of $X$ is denoted by $X^\#=[x_1^\ast,\cdots,x_n^\ast]^\top$. Let $X^\dagger=(X^\#)^\top$.  Clearly, when $n=1$, $X^\dag = X^\ast$. We use  ``$\dag$'' instead of ``$\ast$'' throughout this paper.  $[A,B]=AB-BA$ denotes the commutator between operators $A$ and $B$.  Define two superoperators as
\begin{equation}\begin{aligned}
&\mathrm{Lindbladian}: \mathcal{L}_GX\triangleq -i[X,H]+\mathcal{D}_LX, \\
&\mathrm{Liouvillian}: \mathcal{L}^\star_G\rho \triangleq -i[H,\rho]+\mathcal{D}^\star_L\rho,
\end{aligned}\end{equation}
where $\mathcal{D}_AB=A^\dagger BA-\frac{1}{2}(A^\dagger AB+BA^\dagger A)$ and $\mathcal{D}^\star_AB=ABA^\dagger-\frac{1}{2}(A^\dagger AB+BA^\dagger A)$. We have $\mathrm{Tr}[\rho\mathcal{L}_GX]=\mathrm{Tr}[X\mathcal{L}^\ast_G\rho]$ for a density operator $\rho$ and a bounded operator $X$. Finally, $\otimes$ denotes the tensor product.

\subsection{System and field}\label{subsec:sys and field}

Open Markovian quantum systems can be parameterized conveniently by the $(S,L,H)$ formalism \cite{HP84,GJ09,TNP+11,ZJ12,CKS17,MJ20}. To be specific, for a quantum system driven by  free-propagating Boson fields, $S$ is a unitary operator for example a phase shifter or beamsplitter. The coupling between the system and the fields is described by the operator $L$, and the self-adjoint operator $H$ is the initial system Hamiltonian. $S,L,H$ are all operators on the system Hilbert space  $\mathcal{H}_S$. A free-propagating field is described by its annihilation operator $b(t)$ and creation operator $b^\dagger(t)$ (the adjoint of $b(t)$), which are operators on a Fock space  $\mathcal{H}_F$ and satisfy the following properties
\begin{equation}
b(t)\ket{\Phi_0}=0,~~[b(t),b(r)]=[b^\dagger(t),b^\dagger(r)]=0,~~[b(t),b^\dagger(r)]=\delta(t-r),~~\forall t,r\in \mathbb{R}.
\end{equation}
The integrated annihilation operator and creation operator are defined as $B(t)=\int_{t_0}^{t}b(s)ds$ and $B^\dagger(t)=\int_{t_0}^{t}b^\dagger(s)ds$, respectively, where $t_0$ is the time when the system and field start interaction.

An open quantum system exchanges energy/information with its environment --- the free-propagating Boson fields. Assuming $S=I$ (the identity operator),  the dynamical evolution of the total system (the system of interest plus fields) can be described by a unitary operator $U(t,t_0)$ on the tensor product Hilbert space $\mathcal{H}_S\otimes \mathcal{H}_F$, which is the solution to the following quantum stochastic differential equation (QSDE) \cite{GZ00,BvHJ07,GJ09,GZ15,PZJ16}
\begin{equation}\label{dU}
dU(t,t_0)=\left\{-\left(\frac{1}{2}L^\dagger L+iH\right)dt+LdB^\dagger(t)-L^\dagger dB(t)\right\}U(t,t_0),~~t\geq t_0
\end{equation}
with the initial condition $U(t_0,t_0)=I$.  Let $\ket{\Psi(t)}$ be the state (wavefunction) of the total system at time $t\geq t_0$. Then in the Schr\"{o}dinger picture it is well-known that
\beq\la{schrondiger}
\ket{\Psi(t)}=U(t,t_0)\ket{\Psi(t_0)}.
\eeq
On the other hand, we can also study the dynamics of the system in the Heisenberg picture.  Based on \eqref{dU}, the time evolution of a system operator $X$ on $\mathcal{H}_S$, defined as
\begin{equation}\label{j_t(x)}
j_t(X)\equiv X(t)\triangleq U^\dagger(t)(X\otimes I)U(t),
\end{equation}
is given by \cite{BvHJ07,GJ09,GZ15,PZJ16}
\begin{equation}\label{Jan8}
dj_t(X)=j_t(\mathcal{L}_GX)dt+j_t([L^\dagger,X])dB(t)+j_t([X,L])dB^\dagger(t).
\end{equation}
As  terms $dB(t)$ and $dB^\dagger(t)$ are involved in the time evolution of  $j_t(X)$, it  is an operator on the  tensor product Hilbert space $\mathcal{H}_S\otimes \mathcal{H}_F$. In  this way, the system takes information from the input fields.  After system-field interaction, an output field is generated, which in the input-output formalism is given  by \cite{BvHJ07,GJ09,GZ15,PZJ16}
\begin{equation}\la{B out}
dB_{\mathrm{out}}(t)=L(t)dt+dB(t),
\end{equation}
where $B_{\mathrm{out}}(t)=U^\dagger(I\otimes B(t))U(t)$ denotes the integrated output annihilation operator. Clearly, $B_{\mathrm{out}}(t)$ is an operator on the  tensor product Hilbert space $\mathcal{H}_S\otimes \mathcal{H}_F$. Thus, the output fields carry the system's information which can be measured. More discussions on open quantum systems can be found in, e.g. \cite{WM08,WM10,DP10,AT12,MR15,NY17,ZLWJN17}.

\subsection{Continuous-mode single-photon state}\label{subsec:photon}

A continuous-mode single-photon state in the time domain can be defined as
\begin{equation}\label{spdt}
\ket{\Phi_1}\triangleq B^\dagger(\xi)\ket{\Phi_0},
\end{equation}
where  $\xi(t)$ is the temporal pulse shape which satisfies $\|\xi\|\triangleq\sqrt{\int_{-\infty}^\infty |\xi(t)|^2 dt}=1$, and
\beq \label{B^dag xi}
B^\dagger(\xi)\triangleq\int_{-\infty}^{\infty}\xi(t)b^\dagger(t)dt.
\eeq
Simply speaking,  \eqref{spdt} means that a photon is generated by the creation operator $b^\dagger(t)$ from the vacuum $\ket{\Phi_0}$ with probability $|\xi(t)|^2$, thus the normalization condition $\|\xi\|=1$ guarantees that  exactly one photon is generated. Fourier transforming \eqref{spdt}, yields the continuous-mode single-photon state in the frequency domain,
\begin{equation}\label{spdf}
\ket{\Phi_1}=\int_{-\infty}^{\infty}\xi[i\omega]b^\dagger[i\omega]d\omega\ket{\Phi_0},
\end{equation}
where square brackets are used to indicate that the  designated operators or functions are in the frequency domain. Generally speaking, the continuous-mode single-photon state $\ket{\Phi_1}$ in \eqref{spdf}  describes  a single photon which is coherently superposed over a continuum of frequency modes, with probability amplitudes given by the spectral density function $\xi[i\omega]$. In other words, the probability of finding the photon in the frequency interval $[\omega,\omega+d\omega)$ is $|\xi[i\omega]|^2$, or equivalently,  the probability of finding the photon in the time interval $[t,t+dt)$ is $|\xi(t)|^2$.  More discussions on single-photon states can be found in, e.g., \cite{RL00,FKS10,RWF17,Z20}.

\section{Coherent feedback network}\label{sec:model}

\begin{figure}[htp!]
\centering
\includegraphics[scale=0.4]{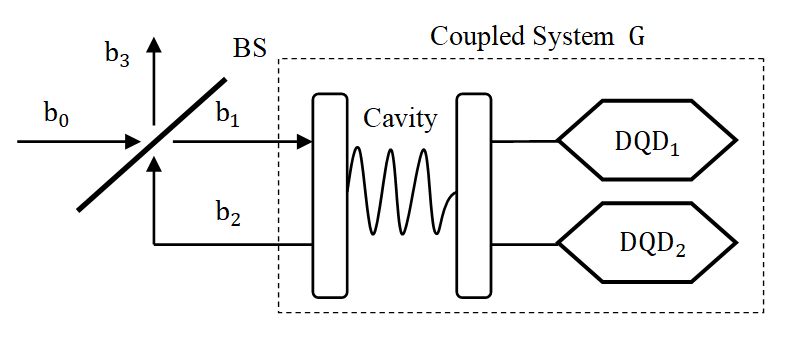}
\caption{Schematic of the coherent feedback network.}
\label{system_Jan16}
\end{figure}

In this paper we focus our sight  on the dynamics of the quantum coherent feedback network as shown in Fig. \ref{system_Jan16}.  In this section, we describe the mathematical model.

 The coupled system $G$ consists of two DQD qubits which are directly coupled to a microwave cavity.  The system $G$ has been recently physically implemented in a semiconductor platform \cite{Van18}. It is worthwhile to notice that the two DQD qubits are separated from each other, thus their interaction is mediated by the microwave cavity which plays the role of a bus. In this paper, we add a  beamsplitter which cascades with $G$, thus forming a coherent feedback network with input  $b_0$ and output $b_3$.

In the $(S,L,H)$ formalism, the beamsplitter $\mathbf{S_b}$ in Fig. \ref{system_Jan16} has  parameters $(S_b,0,0)$ where
\begin{equation}\label{Jansb}
S_b=\left[
      \begin{array}{cc}
        \mu & \sqrt{1-\mu^2} \\
        \sqrt{1-\mu^2} & -\mu \\
      \end{array}
    \right]
\end{equation}
with $\mu$ being the reflection parameter ($0 \leq \mu < 1$). We consider the two DQDs are in the Coulomb-blockade regime with strong intradot and interdot interaction, respectively. Thus, each DQD can be spanned by  two basis states $\ket{0}=\ket{1,0}$ and $\ket{1}=\ket{0,1}$. With furthermore transformation, each DQD has a ground state  $\ket{g}=\alpha\ket{0}-\beta\ket{1}$ and an excited state $\ket{e}= \beta\ket{0}+\alpha\ket{1}$, where $\alpha$ and $\beta$ represent the relationship between the coupling strength and the energy offset.  In this paper, a DQD qubit  has a Hamiltonian of the from
\beq
H_{\mathrm{DQD}} = \frac{\omega_0}{2}\sigma_z,
\eeq
where $\sigma_z=|g\rangle\langle g|-|e\rangle\langle e|$ is a Pauli matrix, and $\omega_0$ is the transition frequency between $\ket{g}$ and $\ket{e}$, which is typically around $5$ GHz \cite{Van18,FLRKS19}. It should be noted that the transition frequency $\omega_0$ is normalized as unit $1$ throughout this paper.  Therefore, the two DQD qubits in Fig. \ref{system_Jan16} are expressed with the $(S,L,H)$ parameters
\begin{equation}
(S_{\mathrm{DQD}_k},L_{\mathrm{DQD}_k},H_{\mathrm{DQD}_k})=(-,-,\frac{1}{2}\delta\omega_k\sigma_{z,k}),~~k=1,2,
\end{equation}
where $\delta\omega_k=\omega_k-\omega_p$ ($k=1,2$) is the detuning between the transition  frequency $\omega_k$ of the $k$th DQD qubit and the carrier frequency $\omega_p$ of the input field. Moreover, $\delta\omega_k>0$ means the red detuning, while the blue detuning is effected by $\delta\omega_k<0$. As the DQD qubits are directly coupled to the cavity, instead of fields, they have no $S$ and $L$ parameters.   Under the rotating-wave approximation (RWA), the Hamiltonian of the coupled system $G$ is
\begin{equation}\label{eq:H}
H_{\mathrm{sys}}=\delta\omega_ra^\dag a+\frac{1}{2}\sum_{k=1}^{2}\delta\omega_k\sigma_{z,k}+\sum_{k=1}^{2}g_k \sin \theta_k(\sigma_{-,k}a^\dag+\sigma_{+,k}a),
\end{equation}
where $\delta\omega_r$ is the frequency detuning between the cavity and the input field, and $\sigma_-=\ket{g}\bra{e}$, $\sigma_+=\ket{e}\bra{g}$ are the lowering and raising operators of the DQD qubit. The third term in \eqref{eq:H} models the direct coupling between the two DQD qubits and the cavity.  (A more detailed description of direct coupling between quantum systems can be found in \cite{ZJ11,ZJ12}). The microwave cavity is lossy, in other words, it exchanges energy with its environment. Let its coupling strength be denoted by $\kappa$.  Then, in the $(S,L,H)$ formalism,  the coupled system $G$ is parameterized as
\begin{equation}\label{Jan17c}
(S_{\mathrm{sys}},L_{\mathrm{sys}},H_{\mathrm{sys}})=\left(1,\sqrt{\kappa}a,\delta\omega_ra^\dag a+\frac{1}{2}\sum_{k=1}^{2}\delta\omega_k\sigma_{z,k}+\sum_{k=1}^{2}g_k \sin \theta_k(\sigma_{-,k}a^\dag+\sigma_{+,k}a)\right).
\end{equation}

\bmrk
The coupled system $G$ has recently been physically implemented on a semiconducting platform \cite{Van18}. The system Hamiltonian $H_{\mathrm{sys}}$ in \eqref{Jan17c} is the same as that in \cite{Van18}, cf. \cite[(C7)]{Van18}. The system $G$ studied in  \cite{Van18} is driven by a laser which is modelled  as a coherent state with amplitude $\alpha$. In this paper, the input is either the vacuum or a single-photon state, in either of these two cases $\alpha=0$. As a result, the coupling operator in \cite[(C13)]{Van18} reduces to $\sqrt{\kappa_{\rm ext}}a$, which is $L_{\mathrm{sys}}$ in \eqref{Jan17c} where $\kappa$ is used instead of $\kappa_{\rm ext}$. Finally, the nonradiative losses and dephasing processes are neglected in our paper. An interested reader may refer to \cite{Van18} for more details of the physics and practical implementation of the  coupled system $G$. Finally, for notational simplicity, we denote $g_k \sin \theta_k$ by $\Gamma_k$ in the remainder of this paper.
\emrk

By means of the $(S,L,H)$ parameters in Eq. \eqref{Jan17c}, the quantum stochastic differential equations for the coupled system $G$ in the Heisenberg picture can be derived as,
\begin{equation}\label{system}
\begin{aligned}
\left[
  \begin{array}{c}
    \dot{\sigma}_{-,1} \\
    \dot{\sigma}_{-,2} \\
    \dot{a} \\
  \end{array}
\right]&=\left[
          \begin{array}{ccc}
            i\delta\omega_1 & 0 & -i\Gamma_1\sigma_{z,1} \\
            0 & i\delta\omega_2 & -i\Gamma_2\sigma_{z,2} \\
            -i\Gamma_1 & -i\Gamma_2 & -i\delta\omega_r-\frac{\kappa}{2} \\
          \end{array}
        \right]\left[
                 \begin{array}{c}
                   \sigma_{-,1} \\
                   \sigma_{-,2} \\
                   a \\
                 \end{array}
               \right]-\left[
                         \begin{array}{c}
                           0 \\
                           0 \\
                           \sqrt{\kappa} \\
                         \end{array}
                       \right]b_1, \\
b_2&=\sqrt{\kappa}a+b_1.
\end{aligned}
\end{equation}
Clearly, \eqref{system} is a bilinear system.  Let the coupled system $G$ be initialized in the state $\ket{g_1}\otimes\ket{g_2}\otimes\ket{0}$; in other words, the two DQD qubits are in their ground states and the cavity is empty. Moreover, let the input field $b_1$ to the coupled system $G$ be in the vacuum state $\ket{\Phi_0}$. Denote
\begin{equation} \label{eq:mar28_1}
|\Phi\rangle=|g_1\rangle\otimes|g_2\rangle\otimes|0\rangle\otimes\ket{\Phi_0},
\end{equation}
and  $X(t)=[
\begin{array}{ccc}
\sigma_{-,1}(t) & \sigma_{-,2}(t) & a(t) \\
\end{array}
]^T$. Then, following the proofs of Lemma 3 and Theorem 5 in \cite{PZJ16}, it can be  shown that
\begin{equation}\label{system1}\begin{aligned}
\langle\Phi|\dot{X}(t)&=A\langle\Phi|X(t)+B\langle\Phi|b_1(t), \\
b_2(t)&=CX(t)+b_1(t),
\end{aligned}\end{equation}
where $B=[
 \begin{array}{ccc}
0 & 0 & -\sqrt{\kappa} \\
\end{array}
]^T$, $C=-B^T$, and
 \begin{equation}\label{Sep22-1}
A = \left[
          \begin{array}{ccc}
            i\delta\omega_1 & 0 & -i\Gamma_1 \\
            0 & i\delta\omega_2 & -i\Gamma_2 \\
            -i\Gamma_1 & -i\Gamma_2 & -i\delta\omega_r-\frac{\kappa}{2} \\
          \end{array}
        \right].
\end{equation}

Next, we look at the closed-loop system. The beamsplitter $\mathbf{S_b}$ in Fig. \ref{system_Jan16} is  a static system,
\beq
\left[
\bey{c}
b_3\\
b_1
\eey
\right]
=S_b
\left[
\bey{c}
b_0\\
b_2
\eey
\right].
\eeq
  By the linear fractional transform \cite[Section 4.4]{ZJ12}, the coherent feedback network in Fig. \ref{system_Jan16} can be expressed in the $(S,L,H)$ formalism as
\begin{equation}\label{Jan17b}
(S_{\mathrm{total}},L_{\mathrm{total}},H_{\mathrm{total}})=\left(1,\sqrt{\frac{1-\mu}{1+\mu}}L_{\mathrm{sys}},H_{\mathrm{sys}}\right).
\end{equation}

\bmrk \label{rem: G and network}
By comparing the $(S,L,H)$ parameters \eqref{Jan17b} of the quantum coherent feedback network and \eqref{Jan17c} of the coupled system $G$, it can be seen that only the coupling operator has been changed by the beamsplitter. More specifically, $\kappa $ is replaced with $\tilde{\kappa}\triangleq\frac{1-\mu}{1+\mu}\kappa$.  Thus, the coupling strength between the coupled system and the input field, or the decay rate of the cavity, can be tuned by changing the beamsplitter reflection parameter $\mu$. Clearly, the coherent feedback network reduces to the open-loop coupled system $G$ when $\mu=0$.
\emrk

By \eqref{Jan17b}, only the coupling strength is changed by the beamsplitter. As a result, similar to \eqref{system1}, the QSDEs for the quantum coherent feedback network acting on $\bra{\Phi}$ is
\begin{equation}\label{Feb29a}
\begin{aligned}
\langle\Phi|\dot{X}(t)&=\tilde{A}\langle\Phi|X(t)+\tilde{B}\langle\Phi|b_0(t), \\
b_3(t)&=\tilde{C} X(t)+b_0(t),
\end{aligned}\end{equation}
where
\begin{equation}\label{Feb29b}
\tilde{A}=\left[
          \begin{array}{ccc}
            i\delta\omega_1 & 0 & -i\Gamma_1 \\
            0 & i\delta\omega_2 & -i\Gamma_2 \\
            -i\Gamma_1 & -i\Gamma_2 & -i\delta\omega_r-\frac{\tilde{\kappa}}{2} \\
          \end{array}
        \right], ~~ \tilde{B}=\left[
          \begin{array}{ccc}
          0 & 0 & -\sqrt{\tilde{\kappa}} \\
          \end{array}
         \right]^T,~~\tilde{C}=-\tilde{B}^T.
\end{equation}
(Recall that  $\tilde{\kappa}=\frac{1-\mu}{1+\mu}\kappa$ as defined in Remark \ref{rem: G and network}.) System \eqref{Feb29a} is of the form of a linear quantum system. By linear systems theory (\cite{ZJ13,NY17,PZJ16}) we get
\begin{equation}\label{Nov28}
\bra{\Phi}b_3(t)=\tilde{C} e^{\tilde{A}(t-t_0)}\bra{\Phi}X(t_0)+\int_{t_0}^{t}g_{\tilde{G}}(t-\tau)\bra{\Phi}b_0(\tau)d\tau,
\end{equation}
where  the impulse response function $g_{\tilde{G}}(t)$ is given by
\begin{equation}\label{Nov28a}
g_{\tilde{G}}(t)=\left\{
\begin{array}{lc}
\delta(t)+\tilde{C}e^{\tilde{A}t}\tilde{B}, & t\geq0, \\
0, & t<0,
\end{array}
\right.
\end{equation}
whose corresponding  transfer function is
\begin{equation}\label{transfer}
\tilde{G}[s]=\frac
{2\Gamma_1^2(s-i\delta\omega_2)+2\Gamma_2^2(s-i\delta\omega_1)+(s-i\delta\omega_1)(s-i\delta\omega_2)(2s-\tilde{\kappa}+2i\delta\omega_r)}
{2\Gamma_1^2(s-i\delta\omega_2)+2\Gamma_2^2(s-i\delta\omega_1)+(s-i\delta\omega_1)(s-i\delta\omega_2)(2s+\tilde{\kappa}+2i\delta\omega_r)}.
\end{equation}
It can be verified that $\tilde{G}[s]$ is an all-pass filter, which only modulates the phase of the input light.  It can be easily seen that the coupled system $G$ is an all-pass filter too.

\section{The steady-state output field state}\label{sec:state}

\subsection{The steady-state output field state}
In this section, assuming the coupled system $G$ is initialized in the state $\ket{g_1}\otimes\ket{g_2}\otimes\ket{0}$ and the input field $b_0$  is in a single-photon state, we aim to derive an analytic expression of the steady-state output field state in the output channel $b_3$. We begin with the following lemma.

\blem\label{lem:A}
All the eigenvalues of the matrix $\tilde{A}$ have non-positive real part.  Moreover, the matrix $\tilde{A}$ is marginally stable if and only if $\delta \omega_1 = \delta \omega_2$.
\elem

{\textbf{Proof.}}
Firstly, as the eigenvalues of the matrix $\tilde{A}+\tilde{A}^\dagger$ are $\left\{0,~0,~-\tilde{\kappa}\right\}$, all the eigenvalues of the matrix $\tilde{A}$ have non-positive real part. In what follows, we show that the matrix $\tilde{A}$ is marginally stable if and only if $\delta \omega_1 = \delta \omega_2$.  Let $\lambda$  be an eigenvalue of the matrix $\tilde{A}$.  Then the characteristic polynomial equation is
\begin{equation}\label{Feb27a}
\lambda^3+(p_1+q_1i)\lambda^2+(p_2+q_2i)\lambda+(p_3+q_3i)=0,
\end{equation} 
where
\begin{equation}\label{Feb27b}
\begin{aligned}
&p_1=\tilde{\kappa}, \\
&q_1=2(\delta\omega_r-\delta\omega_1-\delta\omega_2), \\
&p_2=4\Gamma_1^2+4\Gamma_2^2-4\delta\omega_1\delta\omega_2+4\delta\omega_1\delta\omega_r+4\delta\omega_2\delta\omega_r, \\
&q_2=-2\tilde{\kappa}(\delta\omega_1+\delta\omega_2), \\
&p_3=-4\tilde{\kappa}\delta\omega_1\delta\omega_2, \\
&q_3=-8(\Gamma_1^2\delta\omega_2+\Gamma_2^2\delta\omega_1+\delta\omega_1\delta\omega_2\delta\omega_r).
\end{aligned}\end{equation}
The generalized Hurwitz matrix \cite{SS12} is given by
\begin{equation}\label{Feb27c}
M=\left[
      \begin{array}{cccccc}
        p_1 & q_2i & p_3 & 0 & 0 & 0 \\
        1 & q_1i & p_2 & q_3i & 0 & 0 \\
        0 & p_1 & q_2i & p_3 & 0 & 0 \\
        0 & 1 & q_1i & p_2 & q_3i & 0 \\
        0 & 0 & p_1 & q_2i & p_3 & 0 \\
        0 & 0 & 1 & q_1i & p_2 & q_3i \\
      \end{array}
    \right].
\end{equation}
Let $\Delta_j$ be the $j$-th order determinant of $M$, $j=1,\ldots,6$. Then the Routh like table gives
\begin{equation}\label{Feb27d}
R_1=1,~R_2=p_1,~R_j=\frac{\Delta_{j-1}}{\Delta_{j-2}},~j=3,\ldots,6.
\end{equation}
All the three pairs of points $P_1(R_1,R_2)$, $P_2(R_3,R_4)$, $P_3(R_5,R_6)$ have the same signs ($R_3$ and $R_4$ are pure imaginary) is equivalent to
\begin{equation}\label{Feb27e}
R_1R_2>0,~R_3R_4<0,~R_5R_6>0,
\end{equation}
which is equivalent to $\delta\omega_1\neq\delta\omega_2$. Consequently, according to the Sign Pair Criterion (SPC) \cite{SS12}, all the eigenvalues $\lambda_i$ of system matrix $\tilde{A}$ are in the L.H.S. of the complex plane if and only if $\delta\omega_1\neq\delta\omega_2$. Thus, the matrix $\tilde{A}$ is marginally stable if and only if $\delta \omega_1 = \delta \omega_2$. $\square$

With the aid of Lemma \ref{lem:A}, we are ready to prove the main result of this section.

\bthm\label{theorem3.1}
Let the quantum coherent feedback network be driven by a single-photon input state  \eqref{spdf}, and the coupled system $G$ be initialized in the state $|g_1\rangle\otimes|g_2\rangle\otimes|0\rangle$.  Then the steady-state ($t_0\to -\infty$ and $t\to\infty$) output field state of this quantum coherent feedback network is a single-photon state with  the spectral density function
\begin{equation}\label{fre}
\eta[i\omega]=\tilde{G}[i\omega]\xi[i\omega],
\end{equation}
where the transfer function $\tilde{G}$ is given by \eqref{transfer}.
\ethm

{\textbf{Proof.}}
The proof follows the techniques first developed in \cite{ZJ13} for linear quantum systems and  further generalized  in \cite{PZJ16} to the quantum finite-level systems. The result holds if the following critical conditions hold
\begin{description}
\item[C1.]  $\tilde{C} e^{\tilde{A}(t-t_0)}\bra{\Phi}X(t_0)\to 0$ as $t_0\to-\infty$;
\item[C2.] The transfer function $\tilde{G}[s]$ has no roots on the imaginary axis.
\end{description}

Clearly, if the matrix $\tilde{A}$ is Hurwitz stable, then both {\bf C1} and {\bf C2} are naturally satisfied.  Hence,  by Lemma \ref{lem:A}, it suffices to check the marginal stability case of $\delta \omega_1 = \delta \omega_2$. Let $\delta\omega_1=\delta\omega_2=\delta\omega_s$. Then the system matrix $\tilde{A}$ can be factorized as
\begin{equation}\label{Mar2a}
\tilde{A}=V^{-1}\Lambda_s V,
\end{equation}
where $\Lambda_s=\mathrm{diag}\left\{\lambda_{s1}~,\lambda_{s2},~\lambda_{s3}\right\}$ with the eigenvalues
\begin{equation}\label{Mar2b}\begin{aligned}
&\lambda_{s1}=i\delta\omega_s, \\
&\lambda_{s2}=\frac{1}{4}\left[-\tilde{\kappa}-2i(\delta\omega_r+\omega_s)-\sqrt{(\tilde{\kappa}+2i(\delta\omega_r+\omega_s))^2-16(\Gamma_1^2+\Gamma_2^2)}\right], \\
&\lambda_{s3}=\frac{1}{4}\left[-\tilde{\kappa}-2i(\delta\omega_r+\omega_s)+\sqrt{(\tilde{\kappa}+2i(\delta\omega_r+\omega_s))^2-16(\Gamma_1^2+\Gamma_2^2)}\right].
\end{aligned}\end{equation}
We denote $V=\{v_{ij}\}$ and $V^{-1}=\{w_{ij}\}$ with $i,j=1,2,3$. It can be directly calculated that $v_{13}=0$, $v_{23}=1$, $v_{33}=1$, and $w_{31}=0$. Consequently, we have
\begin{equation}\label{Mar2c}
\tilde{C}e^{\tilde{A}(t-t_0)}\bra{\Phi}X(t_0)=\tilde{C}V^{-1}e^{\Lambda_s(t-t_0)}V\bra{\Phi}X(t_0)\rightarrow0,
\end{equation}
as $t_0\rightarrow-\infty$. Thus, Condition {\bf C1} holds. Moreover, by means of \eqref{Mar2a} and \eqref{Mar2b}, the impulse response function $g_{\tilde{G}}(t)$ in  \eqref{Nov28a} can be calculated as
\begin{equation}\label{Mar2d}\begin{aligned}
g_{\tilde{G}}(t)=\delta(t)+\tilde{C}V^{-1}e^{\Lambda_st}V\tilde{B}=\delta(t)-\tilde{\kappa}\left(w_{32}e^{\lambda_{s2}t}+w_{33}e^{\lambda_{s3}t}\right),~~t\geq0.
\end{aligned}\end{equation}
Consequently, the transfer function $\tilde{G}[s]$ in \eqref{transfer} is of the form
\beq
\tilde{G}[s] = 1-\frac{w_{32}\tilde{\kappa}}{s-\lambda_{s2}} -\frac{w_{33}\tilde{\kappa}}{s-\lambda_{s3}}.
\eeq
As $\lambda_{s2}$ and $\lambda_{s3}$  are both in the L.H.S. of the complex plane, the roots  of $\tilde{G}[s] $ are in the  L.H.S. of the complex plane too.   Thus, Condition {\bf C2} holds.  Consequently, employing the techniques developed in \cite{ZJ13,PZJ16} one can show that the steady-state output field state is  a single-photon state with  pulse shape  given in \eqref{fre}. $\square$

We end this subsection with a final remark.

\bmrk
From the proof of Lemma \ref{lem:A}, it is easy to see that Lemma \ref{lem:A} does not depend on the specific value of $\tilde{\kappa}$; hence it also holds for the matrix $A$. Moreover, when $\mu=0$, the coherent feedback network reduces to the coupled system $G$. As a result, Theorem \ref{theorem3.1} is also true for the coupled system $G$.
\emrk

\subsection{Simulations for the pulse shape of single-photon output state}

In this subsection, we illustrate Theorem  \ref{theorem3.1} by driving the coherent feedback network with a  single photon of the rising exponential pulse shape
\begin{equation}\label{eq:xi}
\xi(t)=\left\{
\begin{array}{cc}
\sqrt{\gamma}e^{\left(\frac{\gamma}{2}+i\omega_p\right)t}, & t\leq0, \\
0, & t>0,
\end{array}
\right.
\end{equation}
where $\omega_p$ is the carrier frequency of the input light field. It can be shown that in the frequency domain this pulse shape has the Lorentzian spectrum and $\gamma$ is the full width at half maximum (FWHM), \cite{RL00,Z20}. In what follows, both the open-loop case (namely the coupled system $G$) and the closed-loop case (namely the coherent feedback network) are simulated.

\begin{figure}[htbp]
\centering
\includegraphics[scale=0.3]{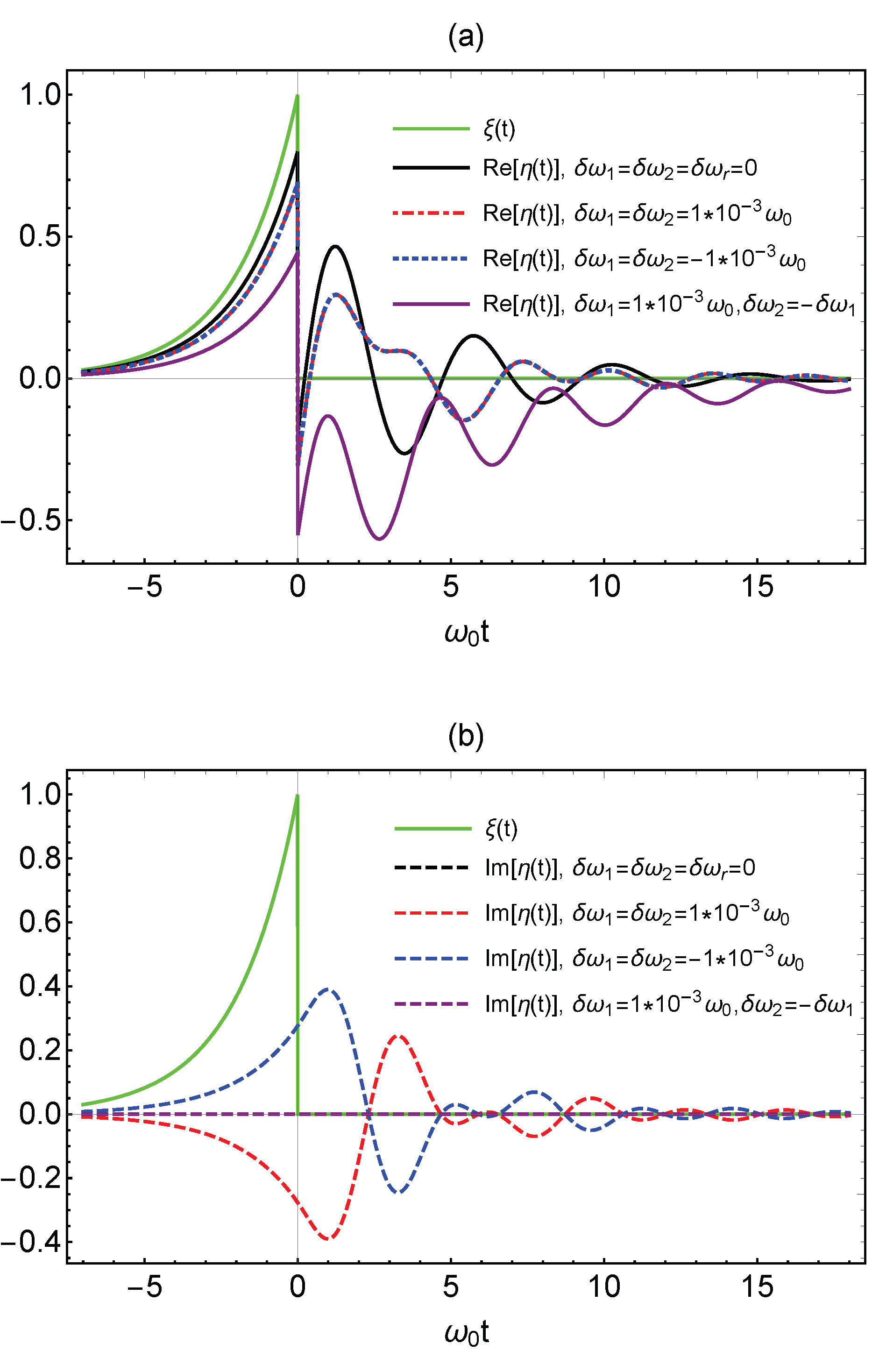}
\caption{The green curve is the pulse shape of the single-photon input state $\xi(t)$. The black solid curve in Fig. \ref{reim0}(a) is the real part of the output pulse shape when the three components (two DQD qubits and the microwave cavity) are tuned into mutual resonance $\delta\omega_1=\delta\omega_2=\delta\omega_r=0$, the corresponding imaginary part is presented in Fig. \ref{reim0}(b) by the black dashed curve. Similar descriptions are applied to the other three cases: red detuning ($\delta\omega_1=\delta\omega_2=1\times10^{-3}\omega_0$), blue detuning ($\delta\omega_1=\delta\omega_2=-1\times10^{-3}\omega_0$), and red+blue detunings ($\delta\omega_1=1\times10^{-3}\omega_0$, $\delta\omega_2=-1\times10^{-3}\omega_0$). The two DQD qubits are equally coupled to the cavity $\Gamma_1=\Gamma_2=1\times10^{-3}\omega_0$. Moreover, $\kappa=\gamma=1\times10^{-3}\omega_0$ and $\delta\omega_r=0$.}
\label{reim0}
\end{figure}

First, we look at the open-loop case. The simulation results are shown in Fig. \ref{reim0}.  We have the following observations:

\begin{itemize}

\item Firstly, it can be observed that the real parts of the output pulse shapes are monotonically increasing for $t<0$.  When $t>0$, they start to oscillate and eventually settle to $0$; see the black solid, red dot-dashed, blue dotted, and purple solid curves in Fig. \ref{reim0}(a). Recall that the pulse shape of a single-photon state is the probability amplitude  of  the single-photon state, which gives the probability of finding the photon. Hence, the oscillating nature of the output pulse shape $\eta(t)$ implies that the photon might be absorbed and emitted by the two DQD qubits from time to time,  thus  giving rise to the quantum Rabi modes. (This is indeed confirmed by the simulations in Subsection \ref{subsec:excitation}.) Rabi oscillations often appear in the Jaynes-Cummings model (interaction between an atom and a resonator or cavity), where the atom alternately emits photons into the resonator and reabsorbs them. It is interesting to see that the Rabi-oscillation-like phenomenon can be observed in the coupled system $G$ driven by a single photon. It is noteworthy to mention that the quantum state collapse and revival phenomena \cite{braumuller2017,Lv18} cannot emerge in this scenario as the photon eventually leaves the system due to the lossy nature of the cavity.

\item Secondly, the pulse shape of the output single photon  for the red detuning case and that for the blue detuning case have the same real parts (see the coincidence between the red dot-dashed curve and the blue dotted curve in Fig. \ref{reim0}(a)); whereas their corresponding imaginary parts are axisymmetric (see the red and blue dashed curves in Fig. \ref{reim0}(b)). Consequently, the phases of the  temporal pulse shapes of the output single photons in these two cases are opposite to each other, which indicates that detunings affect significantly phase-matching. Moreover, this is consistent with the all-pass filter property of the coupled system $G$, in which only the phase of the output single-photon state is changed.

\item Thirdly, when the three components are tuned into mutual resonance  ($\delta\omega_1=\delta\omega_2=\delta\omega_r=0$), the imaginary part of the output pulse shape remains  $0$ all the time (the black dashed curve in Fig. \ref{reim0}(b)). The same result can be found when the detunings for these two DQD qubits with the opposite signs (the purple dashed curve in Fig. \ref{reim0}(b)).  Hence, phase shift can be eliminated in these two scenarios. These phenomena can be verified theoretically. Here we only demonstrate the mutual resonance case. Since the input pulse shape $\xi(t)$ in \eqref{eq:xi} is purely real, it suffices to prove that the impulse response function $g_G(t)$ is a real-valued function. In the resonance case, $g_G(t)$ can be calculated to be
      \begin{equation}\label{Mar2e}
      g_G(t)=\delta(t)-\kappa\left(\cosh\left[\frac{\chi}{4}t\right]-\frac{\kappa}{\chi}\sinh\left[\frac{\chi}{4}t\right]\right) e^{-\frac{\kappa}{4}t},
      \end{equation}
      where $\chi=\sqrt{\kappa^2-16(\Gamma_1^2+\Gamma_2^2)}$. Clearly, no matter whether $\chi$ is real nor not, $g_G(t)$ is always a real-valued function. Therefore, in the case of mutual resonance, the imaginary part of the output pulse shape remains $0$ all the time. The case of red+blue detunings can  be easily verified.

\item Finally, compared with the cases of mutual resonance, red detuning and blue detuning, the  oscillation of the output single photon is more obvious for the case of red+blue detunings (the purple solid curve in Fig. \ref{reim0}(a)), which means that the incident photon escapes from the coupled system much more slowly than that for the other three cases. Since the incident photon escapes from the coupled system much slowly in the case of red+blue detunings, it results in a relatively higher excitation probability for DQD qubits, which is shown in Fig. \ref{Nov3} of Subsection \ref{subsec:excitation}.
\end{itemize}

\begin{figure}[htbp]
\centering
\includegraphics[scale=0.3]{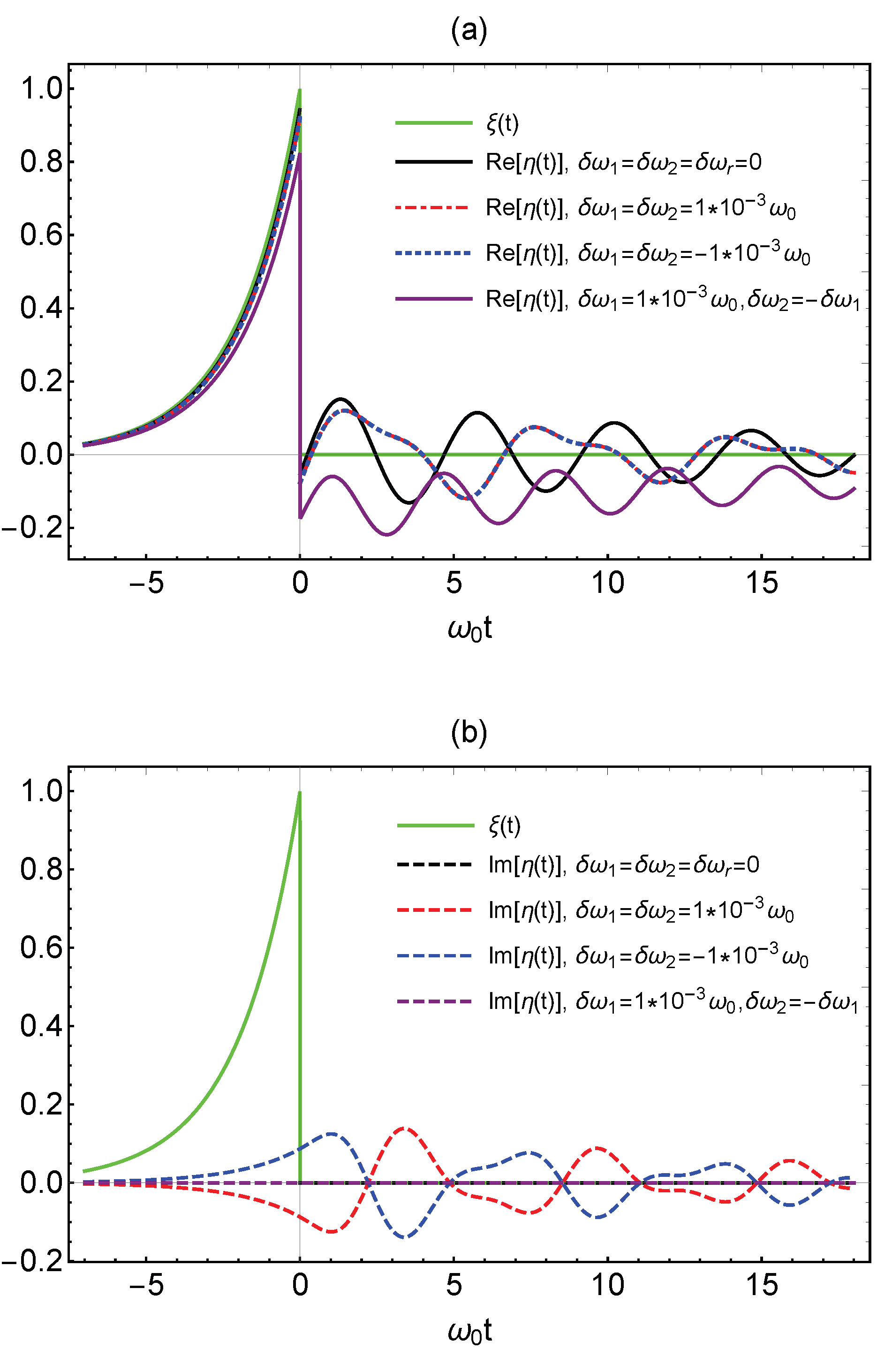}
\caption{The single-photon input pulse shape $\xi(t)$ is plotted with the green curve, while the real and imaginary parts of the output single-photon pulse shapes for the cases of  mutual resonance, red detuning, blue detuning and red+blue detunings are shown as black, red, blue and purple curves in Fig. \ref{reim06}(a) and (b), respectively. The two DQD qubits are equally coupled to the cavity $\Gamma_1=\Gamma_2=1\times10^{-3}\omega_0$. We choose $\kappa=\gamma=1\times10^{-3}\omega_0$, $\delta\omega_r=0$, and $\mu=0.6$.}
\label{reim06}
\end{figure}

Next, we look at the closed-loop case. Choose the beamsplitter reflection parameter $\mu=0.6$. The other parameters are set as those for Fig. \ref{reim0}. The simulation results are shown in Fig. \ref{reim06}.  It can be seen that in contrast to the open-loop case,  the oscillations of both real and imaginary parts of the output pulse shapes in all cases persist for a much longer time; in other words,  coherent feedback elongates the interaction between the system and the input single photon.

\section{The excitation probabilities}\label{sec:excitation}

In Section \ref{sec:state}, we studied how the coherent feedback network in Fig. \ref{system_Jan16} processes a single photon input.  In this section we investigate a closely related problem: how the DQD qubits are excited by an input  single photon? In particular, we compute the excitation probabilities of the first DQD qubit, thus it is named the target  DQD qubit in this section.

\subsection{The excitation probability of the target DQD qubit}\label{subsec:excitation}

In this subsection, we firstly present the\ reduced master equations for the target DQD, then we demonstrate them by some simulators.

 Recall that $\ket{\Phi_1}$ is a single-photon state and $\ket{\Phi_0}$ is the vacuum state of the input field. Denote the expectations
\begin{equation}
\overline{\omega}_t^{mn}(X)=\langle\eta\Phi_m|j_t(X)|\eta\Phi_n\rangle,~~m,n=0,1,
\end{equation}
where $|\eta\rangle$ is the initial state of the coupled system $G$, and by \eqref{Jan8}
\begin{equation}
dj_t(X)=j_t(\mathcal{L}_{\mathrm{total}}X)dt+dB^\dagger(t) j_t([X,L_{\mathrm{total}}])+j_t([L_{\mathrm{total}}^\dagger,X])dB(t)
\end{equation}
with $\mathcal{L}_{\mathrm{total}}X=-i[X,H_{\mathrm{total}}]+L_{\mathrm{total}}^\dagger XL_{\mathrm{total}}-\frac{1}{2}L_{\mathrm{total}}^\dagger L_{\mathrm{total}}X-\frac{1}{2}XL_{\mathrm{total}}^\dagger L_{\mathrm{total}}$.
%
%
%
Define matrices $\overline{\rho}^{mn}(t)$ by means of
\begin{equation} \label{eq:rho}
\mathrm{Tr}[\overline{\rho}^{mn}(t)^\dagger X]\triangleq\overline{\omega}_t^{mn}(X), ~~ m,n=0,1.
\end{equation}

The following result presents the master equations for the quantum coherent feedback network driven by the single-photon state $\ket{\Phi_1}$  in the Schr\"odinger picture.

\blem\cite{GJNC12,Z20}
Master equations for the quantum coherent feedback network driven by the single-photon state $\ket{\Phi_1}$ in \eqref{spdt} in the Schr\"odinger picture are given by
\begin{equation}\label{Oct30d}\begin{aligned}
\dot{\overline{\rho}}^{11}(t)&=\mathcal{L}_{\mathrm{total}}^\star\overline{\rho}^{11}(t)+\xi(t)[\overline{\rho}^{01}(t),L_{\mathrm{total}}^\dagger]+\xi^\ast(t)[L_{\mathrm{total}},\overline{\rho}^{10}(t)], \\
\dot{\overline{\rho}}^{10}(t)&=\mathcal{L}_{\mathrm{total}}^\star\overline{\rho}^{10}(t)+\xi(t)[\overline{\rho}^{00}(t),L_{\mathrm{total}}^\dagger], \\
\dot{\overline{\rho}}^{01}(t)&=\mathcal{L}_{\mathrm{total}}^\star\overline{\rho}^{01}(t)+\xi^\ast(t)[L_{\mathrm{total}},\overline{\rho}^{00}(t)], \\
\dot{\overline{\rho}}^{00}(t)&=\mathcal{L}_{\mathrm{total}}^\star\overline{\rho}^{00}(t),
\end{aligned}\end{equation}
where the initial conditions are
\begin{equation}
\overline{\rho}^{11}(0)=\overline{\rho}^{00}(0)=|\eta\rangle\langle\eta|,~~\overline{\rho}^{10}(0)=\overline{\rho}^{01}(0)=0,
\end{equation}
with $|\eta\rangle$  being the  initial state of the coupled system $G$.
\elem

In what follows, we focus on the density operator of the target DQD qubit  by tracing over the cavity and the 2nd DQD qubit, the reduced density operator for the target DQD qubit is given by
\begin{equation}\label{Oct30}
\rho_{\mathrm{DQD}_1}(t)=\mathrm{Tr}_{\mathrm{DQD}_2}[\mathrm{Tr}_{\mathrm{cav}}[\overline{\rho}(t)]]=\langle g_20|\overline{\rho}(t)|g_20\rangle+\langle g_21|\overline{\rho}(t)|g_21\rangle+\langle e_20|\overline{\rho}(t)|e_20\rangle+\langle e_21|\overline{\rho}(t)|e_21\rangle.
\end{equation}
Let the input single-photon state $\ket{\Phi_1}$ have a Gaussian pulse shape,
\begin{equation}
\xi(t)=\left(\frac{\Omega^2}{2\pi}\right)^{\frac{1}{4}}\exp\left(-\frac{\Omega^2}{4}(t-t_{\rm peak})^2\right),
\end{equation}
where $\Omega$ denotes the photon frequency bandwidth, $t_{\rm peak}$ is the peak arrival time of the photon and  fixed to be $3\omega_0$ in the following simulations. The initial state of the coupled system $G$ is chosen to be $|\eta\rangle=|g_1\rangle\otimes|g_2\rangle\otimes|0\rangle$.  The excitation probabilities of the target DQD qubit are simulated.

\begin{figure}[htp!]
\centering
\includegraphics[scale=0.5]{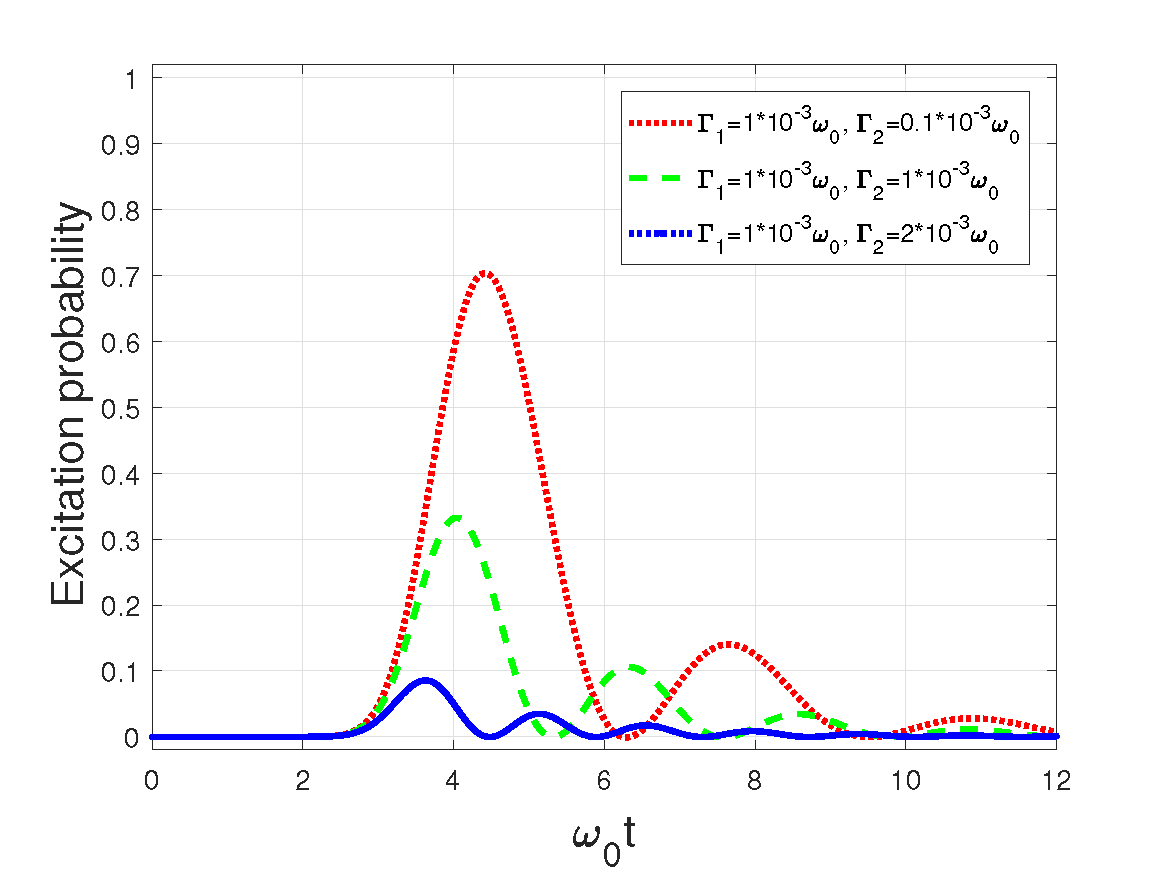}
\caption{The excitation probability of the target DQD qubit with different couplings $\Gamma_2$.  The decay rate of the cavity is $\kappa=1.5\times10^{-3}\omega_0$, the photon frequency bandwidth $\Omega=2.75\kappa$. ($\delta\omega_1=\delta\omega_2=\delta\omega_r=0$ and $\mu=0.2$.) }
\label{exc_coupling}
\end{figure}

\begin{figure}[htp!]
\centering
\includegraphics[scale=0.5]{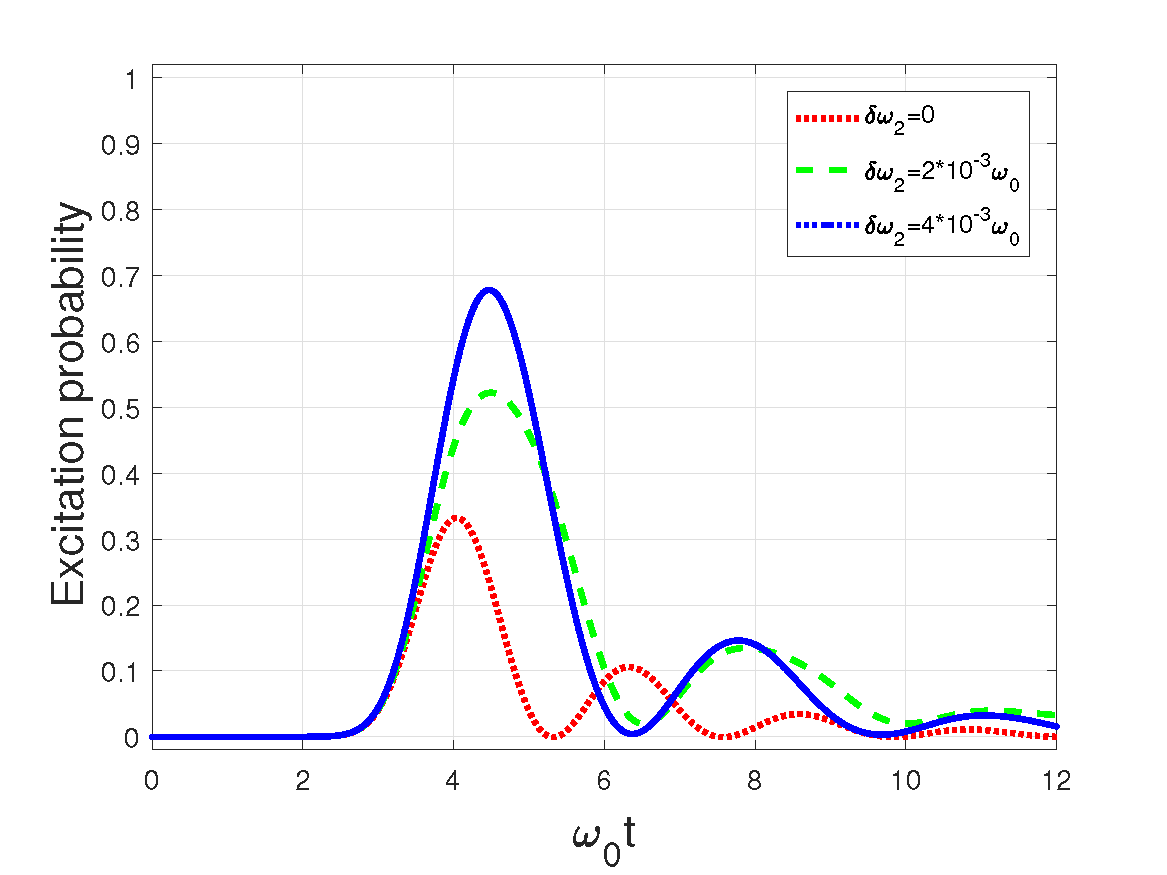}
\caption{The excitation probability of the target DQD qubit with different detunings $\delta\omega_2$. The decay rate of the cavity is $\kappa=1.5\times10^{-3}\omega_0$, the photon frequency bandwidth $\Omega=2.75\kappa$, the couplings are $\Gamma_1=\Gamma_2=1\times10^{-3}\omega_0$, and $\mu=0.2$.}
\label{exc_detuning}
\end{figure}

In Figs. \ref{exc_coupling} and  \ref{exc_detuning}, we plot the excitation probability of the target DQD qubit with different couplings and detunings, respectively. It can be observed that the excitation probability of the target DQD qubit can be significantly improved by reducing the coupling $\Gamma_2$ or increasing the detuning $\delta\omega_2$ of the 2nd DQD qubit, which results in a weak interaction between the cavity and the 2nd DQD qubit.

\begin{figure}[htp!]
\centering
\includegraphics[scale=0.5]{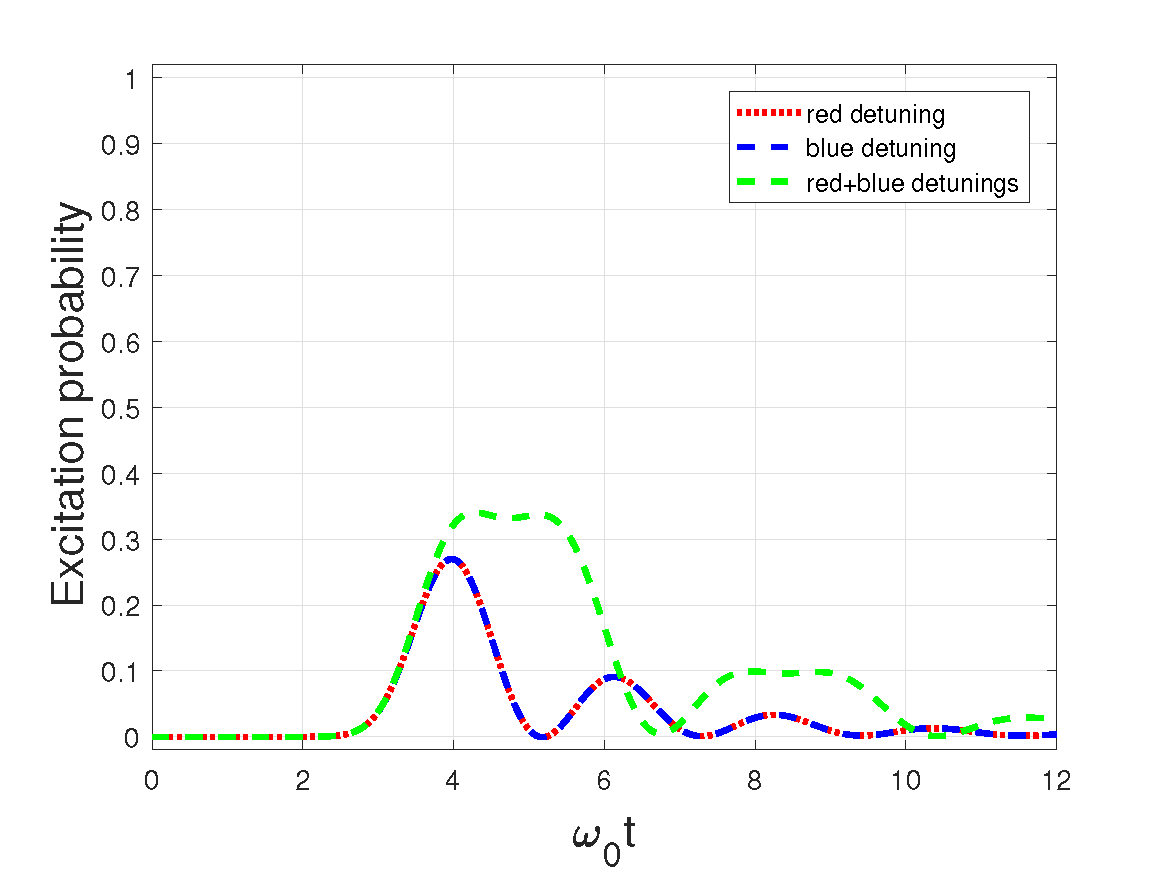}
\caption{The excitation probability of the target DQD qubit with the cases of red detuning, blue detuning and red+blue detunings. The decay rate of the cavity is $\kappa=1.5\times10^{-3}\omega_0$, the photon frequency bandwidth $\Omega=2.75\kappa$, the couplings are $\Gamma_1=\Gamma_2=1\times10^{-3}\omega_0$, and $\mu=0.2$.}
\label{Nov3}
\end{figure}

In Fig. \ref{Nov3}, we plot the excitation probability of the target DQD qubit for three cases of red detuning ($\delta\omega_1=\delta\omega_2=1\times10^{-3}\omega_0$), blue detuning ($\delta\omega_1=\delta\omega_2=-1\times10^{-3}\omega_0$) and red+blue detunings ($\delta\omega_1=1\times10^{-3}\omega_0$, $\delta\omega_2=-1\times10^{-3}\omega_0$). It can be observed that when the detunings for  the two DQD qubits  are with the same sign, the excitation probabilities of the target DQD qubit are identical for the two cases of red detuning and blue detuning. On the other hand, one can increase the excitation probability by setting the two DQD qubits to  red+blue detunings (the green dashed curve), which is consistent with the discussions on the oscillation of the output pulse shapes in Fig. \ref{reim0}.

\begin{figure}[htbp]
\centering
\includegraphics[scale=0.5]{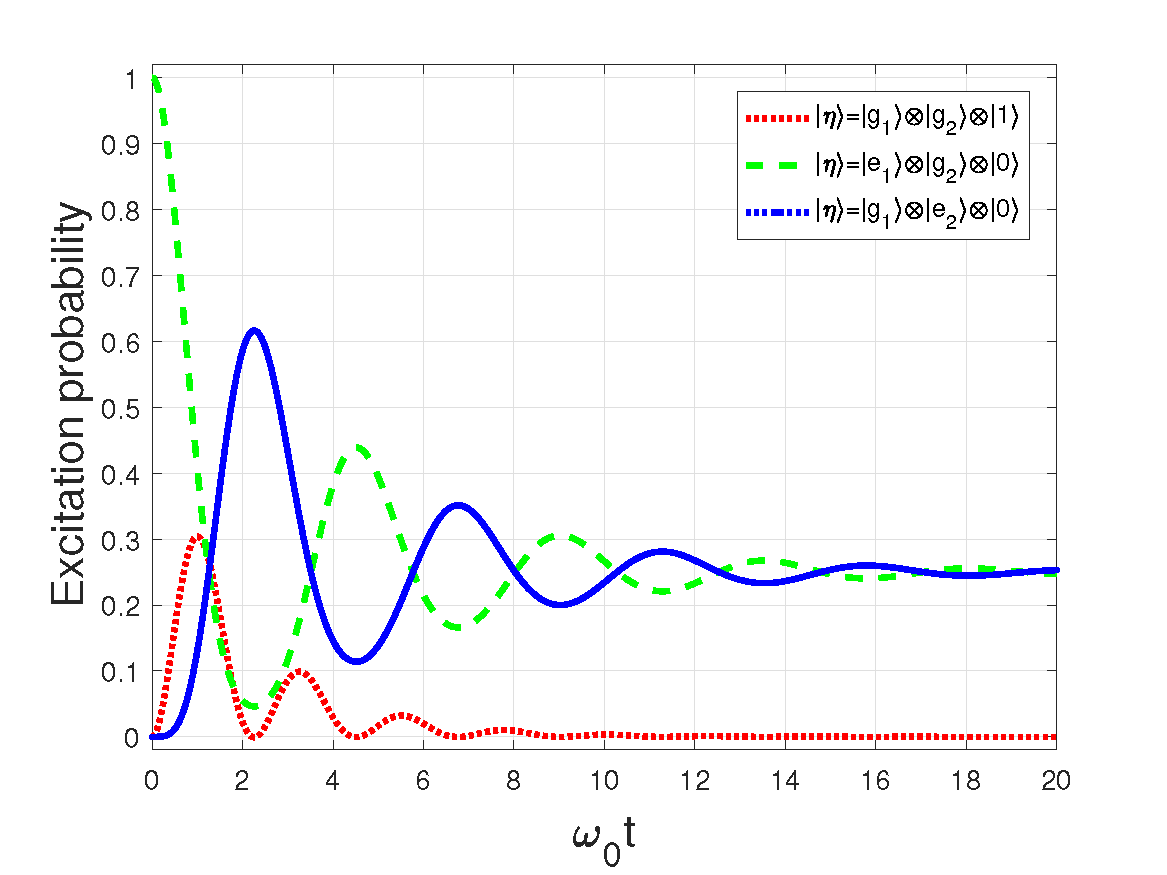}
\caption{The excitation probability of the target DQD qubit with different initial states. We choose $\Gamma_1=\Gamma_2=1\times10^{-3}\omega_0$, $\kappa=1.5\times10^{-3}\omega_0$ and $\mu=0.2$.}
\label{exc_others}
\end{figure}

Finally for comparison, we consider the cases that the coherent feedback network is driven by the vacuum state $\ket{\Phi_0}$ instead of the single-photon state; however, one of the DQD qubits is in its excited state or there is a photon initially in the cavity. The simulation results are shown in Fig. \ref{exc_others} in which all three transitions are tuned into mutual resonance $\delta\omega_1=\delta\omega_2=\delta\omega_r=0$.

In Fig. \ref{exc_others}, it can be observed that the excitation probability of the target DQD qubit approaches $0.25$ when the two DQD qubits are equally coupled to the cavity and the initial state of the coupled system is $|e_1\rangle\otimes|g_2\rangle\otimes|0\rangle$ or $|g_1\rangle\otimes|e_2\rangle\otimes|0\rangle$. This result is consistent with the vacuum Rabi mode splitting with two identical qubits as studied in \cite{Fink09}, where the photon is in each of the DQD qubit with the probability $1/4$. 

\bmrk
It is noteworthy to mention that the discussions above is based on the assumption that all three transitions are tuned into mutual resonance $\delta\omega_1=\delta\omega_2=\delta\omega_r=0$. Not only that, when the two DQD qubits are both equally red- (blue-) detuned with the single-photon input state $\delta\omega_1=\delta\omega_2\neq0$, simulations show that eventually the excitation probability of the target DQD qubit settles to the same value as that for the mutual resonance case. However, when the two DQD qubits are with red+blue detunings $\delta\omega_1=-\delta\omega_2\neq0$, the excitation probability is eventually  $0$, which means that the single photon eventually escapes from the quantum coherent feedback network.
\emrk

\subsection{An analytic result}

The simulation results in Fig. \ref{exc_others} indicate that, when the quantum coherent feedback is driven by a vacuum state while one of the DQD qubits is initially in the excited state, the excitation probability of the target DQD qubit eventually settles to a non-zero value.  In this subsection, we present an analytic result to explain these simulations.

The main result of this subsection is the following theorem.

\bthm\label{alpha}
Assume that the first DQD is initialized in the excited state,  the second DQD is initialized in the ground state, the cavity is initially empty and  the quantum coherent feedback network is driven by the vacuum input state, i.e., the initial joint system-field state is $\ket{e_1g_20\Phi_0}$.   If the central frequencies of the two DQD qubits are not equal (or equivalently $\delta\omega_1\neq\delta\omega_2$), then the steady state of the two DQD qubits is a pure state $|\Phi_{\mathrm{DQD}}(\infty)\rangle=|g_1g_2\rangle$. Otherwise, the steady-state of the joint system-field system is
\begin{equation}\label{Mar20j}
|\Psi(\infty)\rangle=s_1(\infty)  \ket{g_1g_20\Phi_1}+s_2(\infty) |g_1e_20\ket{\Phi_0}+s_3(\infty) |e_1g_20\ket{\Phi_0},
\end{equation}
 where $\ket{\Phi_1}$ is a single-photon state in the output field, and
\begin{equation}\label{Mar20i}
s_1(\infty) = \frac{\alpha}{\sqrt{\alpha^2+1}} , ~~s_2(\infty)=-\frac{\alpha}{\alpha^2+1},~~s_3(\infty)=\frac{1}{\alpha^2+1}
\end{equation}
with $\alpha=\Gamma_1/\Gamma_2$. As a result, the steady state of the two DQD qubits is a mixed state of the form
\begin{eqnarray}
\rho_{\mathrm{DQD}}(\infty)
&=&
|s_1(\infty) |^2|g_1g_2\rangle\langle g_1g_2|+|s_2(\infty) |^2|g_1e_2\rangle\langle g_1e_2|+s_2(\infty) s_3(\infty) ^\ast|g_1e_2\rangle\langle e_1g_2|
\nonumber
\\
&&+s_3(\infty) s_2(\infty) ^\ast|e_1g_2\rangle\langle g_1e_2|+
|s_3(\infty) |^2|e_1g_2\rangle\langle e_1g_2|.
\label{Nov16}
\end{eqnarray}
\ethm

{\textbf{Proof.}}
As there is only one excitation,  the joint system-field state $\ket{\Psi(t)}$ is of the form
\begin{equation}\label{Mar20a}
\ket{\Psi(t)}=s_1(t)\int_{0}^{t}\eta(\tau)b^\dagger(\tau)d\tau|g_1g_20\ket{\Phi_0}+s_2(t)|g_1e_20\ket{\Phi_0}+s_3(t)|e_1g_20\ket{\Phi_0}+s_4(t)|g_1g_21\ket{\Phi_0},
\end{equation}
where the initial condition is $s_2(0)=s_4(0)=0$, $s_3(0)=1$.   $\eta(\tau)$ denotes the pulse shape of the single-photon output state with $\|\eta\|=1$. As the integral is 0 at $t=0$, it is not necessary to impose an initial condition on $s_1(t)$. Thus, we just assume $s_1(0)=1$. The normalization condition for $\ket{\Psi(t)}$ is
\begin{equation}\label{Mar20e}
|s_1(t)|^2\int_{0}^{t}|\eta(\tau)|^2d\tau+\sum_{j=2}^{4}|s_j(t)|^2=1,
\end{equation}
and in the stationary state ($t\rightarrow\infty$)
\begin{equation}\label{Mar20f}
\sum_{j=1}^{4}|s_j(\infty)|^2=1.
\end{equation}
By \eqref{dU}-\eqref{schrondiger} and \cite[Eq. (11.2.18)]{GZ00}, we have the following It\={o} quantum stochastic Schr\"{o}dinger equation (QSSE) 
\begin{equation}\label{Mar20b}\begin{aligned}
d\ket{\Psi(t)}=&\left\{-\left(\frac{1}{2}L_{\rm sys}^\dagger L_{\rm sys}+iH_{\rm sys}\right)dt+LdB^\dagger(t)\right\}\ket{\Psi(t)}
\\
&=\bigg\{\bigg[-i\delta\omega_r-\frac{i}{2}\delta\omega_1-\frac{i}{2}\delta\omega_2-\frac{\tilde{\kappa}}{2}\bigg]s_4(t)-i\Gamma_1s_3(t)
-i\Gamma_2s_2(t)\bigg\}|g_1g_21\ket{\Phi_0} dt 
\\
&+\bigg[-\frac{i}{2}(\delta\omega_1-\delta\omega_2)s_2(t)-i\Gamma_2s_4(t)\bigg]|g_1e_20\ket{\Phi_0} dt \\
&+\bigg[\frac{i}{2}(\delta\omega_1-\delta\omega_2)s_3(t)-i\Gamma_1s_4(t)\bigg]|e_1g_20\ket{\Phi_0} dt \\
&+\bigg[-\frac{i}{2}(\delta\omega_1+\delta\omega_2)s_1(t)\int_{0}^{t}\eta(\tau)b^\dagger(\tau)d\tau\bigg]|g_1g_20\ket{\Phi_0} dt+\sqrt{\tilde{\kappa}}s_4(t)dB^\dagger(t)|g_1g_20\ket{\Phi_0}.
\end{aligned}\end{equation}
On the other hand, differentiating \eqref{Mar20a} with respect to $t$, yields
\begin{equation}\label{Mar20c}\begin{aligned}
d|\Psi(t)\rangle=&\dot{s}_4(t)|g_1g_21\ket{\Phi_0} dt+\dot{s}_2(t)|g_1e_20\ket{\Phi_0} dt+\dot{s}_3(t)|e_1g_20\ket{\Phi_0} dt \\
&+\bigg[\dot{s}_1(t)\int_{0}^{t}\eta(\tau)b^\dagger(\tau)d\tau\bigg]|g_1g_20\ket{\Phi_0} dt+s_1(t)\eta(t)dB^\dagger(t)|g_1g_20\ket{\Phi_0}.
\end{aligned}\end{equation}
By comparing the first $3$ terms in \eqref{Mar20b} and \eqref{Mar20c}, one  gets that
\begin{equation}\label{Mar20d}
\left\{\begin{aligned}
&\dot{s}_2(t)=-\frac{i}{2}(\delta\omega_1-\delta\omega_2)s_2(t)-i\Gamma_2s_4(t), \\
&\dot{s}_3(t)=\frac{i}{2}(\delta\omega_1-\delta\omega_2)s_3(t)-i\Gamma_1s_4(t), \\
&\dot{s}_4(t)=\bigg[-i\delta\omega_r-\frac{i}{2}\delta\omega_1-\frac{i}{2}\delta\omega_2-\frac{\tilde{\kappa}}{2}\bigg]s_4(t)-i\Gamma_1s_3(t)
-i\Gamma_2s_2(t)
\end{aligned}
\right.
\end{equation}
with the initial condition  $[s_2(0) ~~ s_3(0) ~~ s_4(0)]=[ 0 ~~ 1~~0]$.

In what follows, we find the stationary solution of  \eqref{Mar20d}   by sending $t\rightarrow\infty$.

If $\delta\omega_1\neq\delta\omega_2$, it can be readily shown that the stationary solution is
\begin{equation}\label{Mar20g}
s_2(\infty)=s_3(\infty)=s_4(\infty)=0,~~s_1(\infty)=1.
\end{equation}
Thus, in this case, the stationary state of the coherent feedback network is
\begin{equation}\label{Mar20h}
|\Psi(\infty)\rangle=\int_{0}^{\infty}\eta(\tau)b^\dagger(\tau)d\tau|g_1g_20\ket{\Phi_0} \equiv \ket{g_1g_2 0\Phi_1},
\end{equation}
where $\ket{\Phi_1} \triangleq \int_{0}^{\infty}\eta(\tau)b^\dagger(\tau)d\tau\ket{\Phi_0}$ is a single-photon state. In other words, eventually the two DQD qubits are in the ground state, the cavity is empty, and the output field contains  a single photon.

If $\delta\omega_1=\delta\omega_2$, substituting $\Gamma_1=\alpha \Gamma_2$ into \eqref{Mar20d} and solving it under the initial condition  $[s_2(0) ~~ s_3(0) ~~ s_4(0)]=[ 0 ~~ 1~~0]$, we get
\begin{equation}\label{Mar22b}
\begin{aligned}
s_2(t)&=\frac{\Gamma_1\Gamma_2}{2(\Gamma_1^2+\Gamma_2^2)\lambda_3^\prime}
\bigg[(-e^{\lambda_1^\prime t}+e^{\lambda_2^\prime t})(\tilde{\kappa}+2i\delta\omega_2+2i\delta\omega_r)+\lambda_3^\prime e^{\lambda_1^\prime t}+\lambda_3^\prime e^{\lambda_2^\prime t}-2\lambda_3^\prime\bigg], \\
s_3(t)&=\frac{1}{2(\Gamma_1^2+\Gamma_2^2)\lambda_3^\prime}
\bigg[2\Gamma_2^2\lambda_3^\prime+4\lambda_2^\prime \Gamma_1^2 e^{\lambda_1^\prime t}-4\lambda_1^\prime \Gamma_1^2 e^{\lambda_2^\prime t}\bigg], \\
s_4(t)&=\frac{2i\Gamma_1(-e^{\lambda_1^\prime t}+e^{\lambda_2^\prime t})}{\lambda_3^\prime},
\end{aligned}
\end{equation}
where
\begin{equation}\label{Mar22c}\begin{aligned}
\lambda_1^\prime&=\frac{1}{4}\left[-\tilde{\kappa}-2i(\delta\omega_2+\delta\omega_r)
-\sqrt{-16(\Gamma_1^2+\Gamma_2^2)+(\tilde{\kappa}+2i\delta\omega_2+2i\delta\omega_r)^2}\right], \\
\lambda_2^\prime&=\frac{1}{4}\left[-\tilde{\kappa}-2i(\delta\omega_2+\delta\omega_r)
+\sqrt{-16(\Gamma_1^2+\Gamma_2^2)+(\tilde{\kappa}+2i\delta\omega_2+2i\delta\omega_r)^2}\right], \\
\lambda_3^\prime&=\sqrt{-16(\Gamma_1^2+\Gamma_2^2)+(\tilde{\kappa}+2i\delta\omega_2+2i\delta\omega_r)^2}.
\end{aligned}\end{equation}
Sending $t\to \infty$, we obtain
\begin{equation}\label{Mar20ii}
s_2(\infty)=-\frac{\alpha}{\alpha^2+1},~~s_3(\infty)=\frac{1}{\alpha^2+1},~~s_4(\infty)=0.
\end{equation}
By the normalization condition \eqref{Mar20f}, we have $|s_1(\infty)|^2=\frac{\alpha^2}{\alpha^2+1}$. Furthermore, if we let the phase of $s_1(\infty)$ be absorbed by the pulse shape $\eta$, then  $s_1(\infty)$ is a positive number $\frac{\alpha}{\sqrt{\alpha^2+1}}$.  In this case, the stationary state of the whole system is Eq. \eqref{Mar20j}.
 Finally, by tracing over the cavity and the field, the steady state of the two DQD qubits is given by
\begin{equation}\label{Mar20k}
\rho_{\mathrm{DQD}}=\langle0\Psi_1|\Psi(t)\rangle\langle\Psi(t)|0\Phi_1\rangle+\langle0\Phi_0|\Psi(t)\rangle\langle\Psi(t)|0\ket{\Phi_0}.
\end{equation}
Substituting \eqref{Mar20j} into \eqref{Mar20k}, yields \eqref{Nov16}. $\square$

\bmrk
When the two DQD qubits have the same central frequency, by Theorem \ref{alpha} we see that in the steady state the output field and the two DQD qubits are in a superposition pure state \eqref {Mar20j}, which means that a photon exists simultaneously in the output field and inside the 2-DQD qubit system.  If $\alpha=0$, i.e., the first DQD qubit is decoupled from the other components. In this case, the whole system is in the state $ |e_1g_20\ket{\Phi_0}$. This is confirmed by \eqref{Nov16}. On the other hand, if $\alpha=\infty$, then the second DQD qubit is decoupled. In this case $s_1(\infty)=1$ and $s_2(\infty) =s_3(\infty)=0$ and Eq. \eqref {Mar20j} reduces to $\ket{g_1g_20\Phi_1}$, in other words, the output field contains a single photon,  both DQD qubits are in the ground state and the cavity is empty.
\emrk

\bmrk
By comparing the last two terms in \eqref{Mar20b} and \eqref{Mar20c}, one  gets that
\beqn
\dot{s}_1(t) &=&  -\frac{i}{2}(\delta\omega_1+\delta\omega_2)s_1(t),
\label{eq:mar29_temp1a}
\\
s_1(t)\eta(t)&=&\sqrt{\tilde{\kappa}}s_4(t),
\label{eq:mar29_temp1b}
\eeqn
with the initial conditions $s_1(0)=1$ and $s_4(0)=0$. As the system of equations \eqref{Mar20d} is linear, $s_4(t)$ can be obtained. Moreover, $s_1(t)$ can be obtained from \eqref{eq:mar29_temp1a}. As a result, the pulse shape $\eta(t)$ in both cases ($\delta\omega_1=\delta\omega_2$ and  $\delta\omega_1\neq\delta\omega_2$)  can be calculated.
\emrk

\bmrk
Comparing the expression \eqref{Nov16} with the bright states $|\pm\rangle_{r3}$ for the coherent input case in \cite[Fig. 2]{Van18},  each DQD qubit being in its excited state shares the same ratio of probability $\Gamma_1^2/\Gamma_2^2=\alpha^2$. However, in contrast to the fixed probability $1/2$ in \cite[Fig. 2]{Van18}, the probability of the two coupled DQD qubits in the ground state $|g_1g_2\rangle$ is controllable with the value $\frac{\alpha^2}{\alpha^2+1}$, which is also the probability of the single-photon escaping from the coupled system. By Theorem \ref{alpha}, it can be seen that the steady state of the two coupled DQD qubits is independent of the cavity decay rate $\kappa$ and beamsplitter parameter $\mu$ when the quantum coherent feedback network is driven by vacuum input.
\emrk


\section{Full inversion of the target DQD qubit}\label{sec:inverse}

The discussions in Subsection \ref{subsec:excitation} tell us that the target DQD qubit cannot be fully excited by the single-photon input state with a Gaussian pulse shape. In this section, we study whether there is a desired input pulse shape with which the input single photon is able to fully excite the target DQD qubit.  For simplicity, assume that all three transitions are tuned into mutual resonance, i.e., $\delta\omega_1=\delta\omega_2=\delta\omega_r=0$.

The following lemma can be proved in the similar way as that for \cite[Lemma 3]{PZJ16}.

\blem\label{lem:Phi}
Let the initial state for the coherent feedback network be   $|\Phi\rangle=|g_1\rangle\otimes|g_2\rangle\otimes|0\rangle\otimes\ket{\Phi_0}$ in \eqref{eq:mar28_1}. Then
\beq
U(t,t_0)|\Phi\rangle  = e^{i\alpha_t}|\Phi\rangle.
\eeq
\elem

As $ e^{i\alpha_t}$ only adds a global phase to the state of the whole system, we ignored it in the following discussions.

 The following theorem is the main result of this section, it gives an analytic form of the pulse shape of the single-photon input state, by which the input single photon can fully excite the target DQD qubit at a prescribed time.

\bthm\label{thm:full inversion}
Assume that the two DQD qubits are initialized in the ground state and  the cavity is empty, i.e., the initial state of the coupled system $G$ is $|\eta\rangle=|g_1\rangle\otimes|g_2\rangle\otimes|0\rangle$. If the pulse shape of the single-photon input state is
\begin{equation}\label{Sep22-3}
\begin{aligned}
\tilde{\nu}_k(r)=\frac{2i\Gamma_k\sqrt{\tilde{\kappa}}}{\sqrt{\tilde{\kappa}^2-16\Gamma_k^2}}\left(e^{\frac{-\tilde{\kappa}-\sqrt{\tilde{\kappa}^2-16\Gamma_k^2}}{4}(t-r)}-
e^{\frac{-\tilde{\kappa}+\sqrt{\tilde{\kappa}^2-16\Gamma_k^2}}{4}(t-r)}\right),~~r\leq t,k=1,2,
\end{aligned}\end{equation}
where $t$ is the terminal time of the single-photon input pulse shape, then the $k$-th DQD qubit can be transferred from the ground state $|g_k\rangle$ to the excited state $|e_k\rangle$ at the controllable time $t$. Moreover, in this case, the other DQD qubit must be decoupled from the cavity.
\ethm

{\textbf{Proof.}}
 Conjugating both sides of \eqref{Feb29a} yields
\begin{equation}\label{Sep6}
\dot{X}^\dagger(t)|\Phi\rangle=X^\dagger(t)|\Phi\rangle \tilde{A}^{\dagger}+ b^\dagger(t)|\Phi\rangle {\tilde{B}}^T,
\end{equation}
whose solution is
\begin{equation}\label{Sep6-1}
\begin{aligned}
X^\dagger(t)|\Phi\rangle=&\left[
  \begin{array}{ccc}
    \sigma_{+,1}(t)|\Phi\rangle & \sigma_{+,2}(t)|\Phi\rangle & a^\dagger(t)|\Phi\rangle \\
  \end{array}
\right] \\
=&X^\dagger(t_0)e^{\tilde{A}^\dagger(t-t_0)}|\Phi\rangle+\int_{t_0}^{t}{\tilde{B}}^T e^{\tilde{A}^\dagger(t-r)} b^\dagger(r)|\Phi\rangle dr.
\end{aligned}\end{equation}
%
Define exponentially rising  functions
\begin{equation}\label{Sep22-2}
\left[
  \begin{array}{c}
    \tilde{\nu}_1(r) \\
    \tilde{\nu}_2(r) \\
    \tilde{\nu}_3(r) \\
  \end{array}
\right]^T={\tilde{B}}^T e^{\tilde{A}^\dagger(t-r)},~~r\leq t.
\end{equation}
Then define
\begin{equation}
\left[
  \begin{array}{c}
    \nu_1(r) \\
    \nu_2(r) \\
    \nu_3(r) \\
  \end{array}
\right]=\left[
          \begin{array}{c}
            \tilde{\nu}_1(r)\Theta(t-r) \\
            \tilde{\nu}_1(r)\Theta(t-r) \\
            \tilde{\nu}_1(r)\Theta(t-r) \\
          \end{array}
        \right],
\end{equation}
where $\Theta(s)$ is the Heaviside function. Similar to the proof of Theorem \ref{theorem3.1},  by sending $t_0\rightarrow-\infty$, the solution \eqref{Sep6-1} can be derived as
\begin{equation}\label{eq:mar24_1}
\begin{aligned}
&\left[
  \begin{array}{ccc}
    \sigma_{+,1}(t)|\Phi\rangle & \sigma_{+,2}(t)|\Phi\rangle & a^\dagger(t)|\Phi\rangle \\
  \end{array}
\right] \\
=&\int_{-\infty}^{\infty}\left[
                                 \begin{array}{ccc}
                                   \nu_1(r) & \nu_2(r) & \nu_3(r) \\
                                 \end{array}
                               \right]
 b^\dagger(r)|\Phi\rangle dr \\
= &\left[
                                             \begin{array}{ccc}
                                               B^\dagger(\nu_1)|\Phi\rangle & B^\dagger(\nu_2)|\Phi\rangle & B^\dagger(\nu_3)|\Phi\rangle \\
                                             \end{array}
                                           \right],
\end{aligned}
\end{equation}
where $B^\dagger(\nu_i)$ is the creation operator as defined in \eqref{B^dag xi}, which can generate a single-photon state with the pulse shape $\nu_i(t)$, $i=1,2,3$. Particularly,
\begin{equation}\label{Oct30b}
\sigma_{+,1}(t)|\Phi\rangle=B^\dagger(\nu_1)|\Phi\rangle,~~\sigma_{+,2}(t)|\Phi\rangle=B^\dagger(\nu_2)|\Phi\rangle.
\end{equation}
Without loss of generality, let $k=1$. Let the single-photon input state be
\begin{equation}\label{eq:mar25_1}
\ket{\Phi_1}=B^\dagger(\nu_1)\ket{\Phi_0} =B^\dagger(\tilde{\nu}_1)\ket{\Phi_0}.
\end{equation}
By the Schr\"{o}dinger equation \eqref{schrondiger},  the  state of the quantum coherent feedback network at time $t$ can be calculated as
\begin{equation}\label{Sep6-2}\begin{aligned}
\ket{\Psi(t)}&=U(t,t_0)|g_1\rangle\otimes|g_2\rangle\otimes|0\rangle\otimes\ket{\Phi_1} \\
&=U(t,t_0)B^\dagger(\nu_1)|g_1\rangle\otimes|g_2\rangle\otimes|0\rangle\otimes\ket{\Phi_0} \\
&=U(t,t_0)\sigma_{+,1}(t)|\Phi\rangle \\
&=U(t,t_0)\sigma_{+,1}(t)U^\dagger(t,t_0)U(t,t_0)|\Phi\rangle \\
&=U(t,t_0)\sigma_{+,1}(t)U^\dagger(t,t_0)|\Phi\rangle \\
&=\sigma_{+,1}(t_0)|\Phi\rangle \\
&=|e_1\rangle\otimes|g_2\rangle\otimes|0\rangle\otimes\ket{\Phi_0},
\end{aligned}\end{equation}
where \eqref{Oct30b} has been used in the 3rd step and Lemma \ref{lem:Phi} is used in the 4th step. By the last step of \eqref{Sep6-2}, it is obvious that the 1st DQD is fully excited. In what follows, we derive the explicit expression of the input photon pulse shape, which transfers the state of the 1st DQD qubit at any controllable time $t$. Actually, by \eqref{Feb29b} and \eqref{Sep22-2}, one can obtain
\begin{equation}\label{Oct30c}\begin{aligned}
\tilde{\nu}_1(r)=\frac{2i\Gamma_1\sqrt{\tilde{\kappa}}}{\sqrt{\tilde{\kappa}^2-16g_c^2}}\left(e^{\frac{-\tilde{\kappa}-\sqrt{\tilde{\kappa}^2-16g_c^2}}{4}(t-r)}-
e^{\frac{-\tilde{\kappa}+\sqrt{\tilde{\kappa}^2-16g_c^2}}{4}(t-r)}\right),~~r\leq t,
\end{aligned}\end{equation}
where $g_c^2=\Gamma_1^2+\Gamma_2^2$.
  It should be noted that as $\ket{\Phi_1} = B^\dag(\tilde{\nu}_1)\ket{\Phi_0}$ is a single-photon state,  the input pulse shape $\tilde{\nu}_1(r)$ must satisfy the normalization condition
\begin{equation}\label{Sep22-4}
\int_{-\infty}^{t}|\tilde{\nu}_1(r)|^2dr=1,
\end{equation}
which in turn yields $g_1^2/g_c^2=1$. Thus, the pulse shape $\tilde{\nu}_1(r)$ in \eqref{Oct30c}  is exactly that in \eqref{Sep22-3} for $k=1$. Moreover, in this case, we have $\Gamma_2=0$, in other words, the 2nd DQD qubit must be decoupled from the cavity. Similar derivations can be applied to the case of the 2nd DQD qubit. $\square$

\begin{figure}[htp!]
\centering
\includegraphics[scale=0.6]{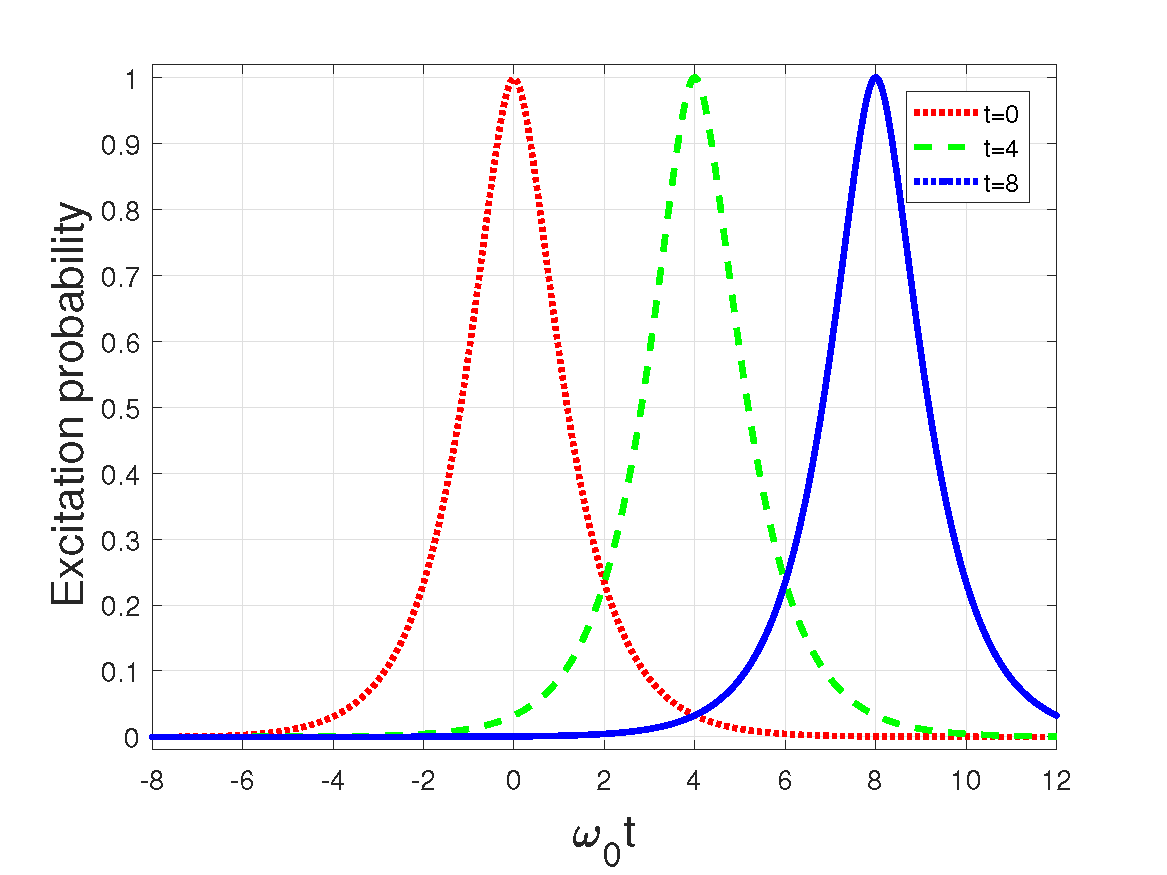}
\caption{The excitation probability of the 1st DQD qubit for different terminal time $t$. ($\mu=0.2$, $\kappa=7.5\times10^{-3}\omega_0$, $\Gamma_1=1\times10^{-3}\omega_0$, and $\Gamma_2=0$.) The terminal time of the single-photon input pulse shape, also known as the controllable time for the DQD qubit excitation, is given by $t=0$, $t=4$, and $t=8$, respectively.}
\label{inverse}
\end{figure}

\begin{figure}[htp!]
\centering
\includegraphics[scale=0.6]{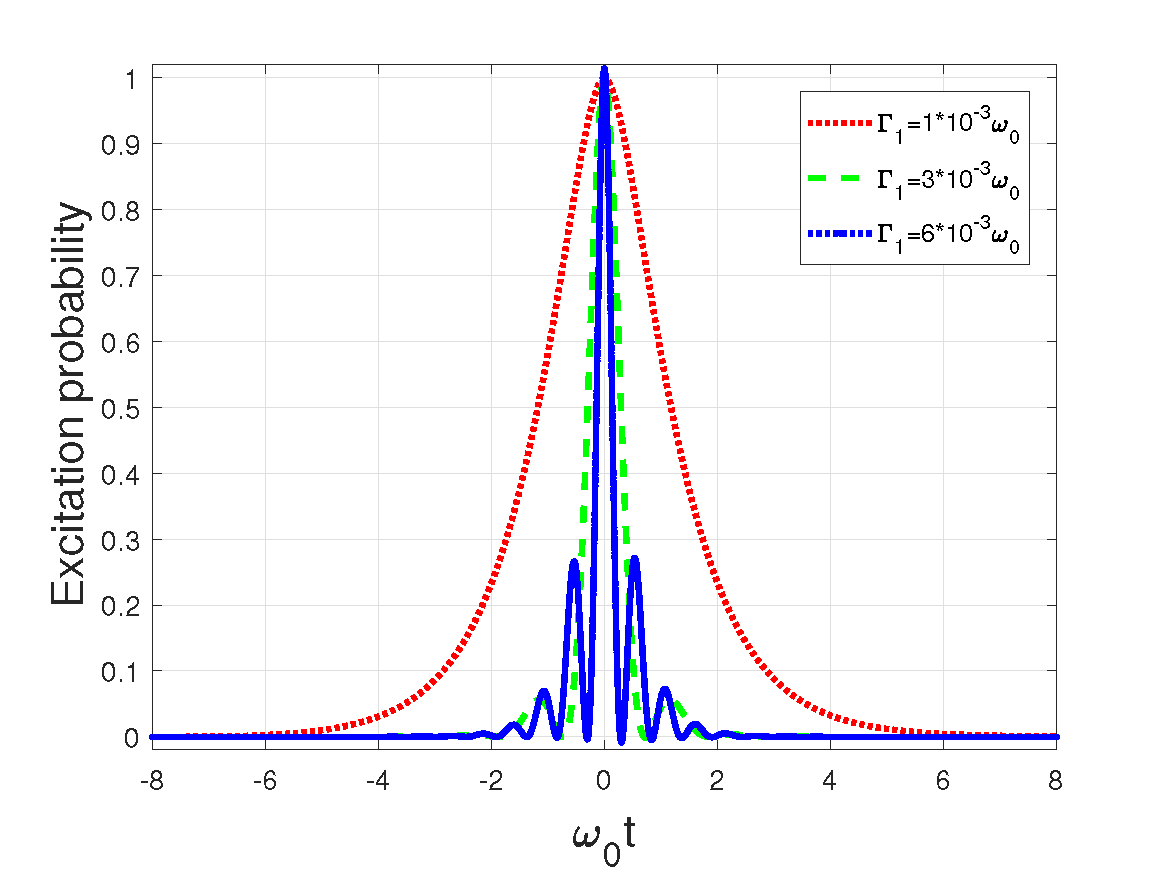}
\caption{The excitation probability of the 1st DQD qubit under different $\Gamma_1$. We fix the terminal time $t=0$ and choose $\mu=0.2$, $\kappa=7.5\times10^{-3}\omega_0$, and $\Gamma_2=0$. The coupling between the 1st DQD qubit and the cavity is chosen to be $\Gamma_1=1\times10^{-3}\omega_0$, $\Gamma_1=3\times10^{-3}\omega_0$, and $\Gamma_1=6\times10^{-3}\omega_0$, respectively.}
\label{inverse_rabi}
\end{figure}

\begin{figure}[htp!]
\centering
\includegraphics[scale=0.8]{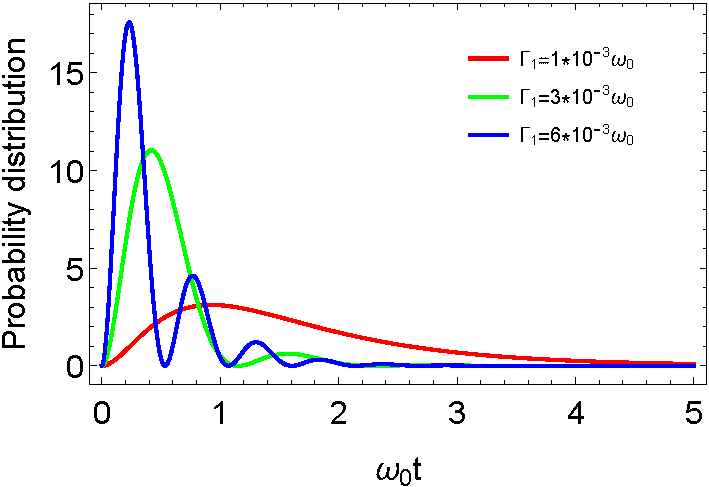}
\caption{The probability distributions for the output single-photon states with respect to the three cases discussed in Fig. \ref{inverse_rabi}, in which Rabi oscillation phenomena occur when the decay rate of the cavity and the coupling satisfy $\tilde{\kappa}<4\Gamma_1$.}
\label{dec8}
\end{figure}

In Fig. \ref{inverse}, the excitation probabilities of the 1st DQD qubit with different terminal times of the single-photon input pulse shape are plotted, which demonstrate that the pulse shape \eqref{Sep22-3} of the single-photon input state can be used to fully excite the target DQD qubit at a controllable time $t$. Interestingly, when the decay rate $\kappa$ of the cavity  and the coupling $\Gamma_1$ satisfy $\tilde{\kappa}<4\Gamma_1$, the matrix $\tilde{A}$ has a pair of complex eigenvalues, thus the Rabi oscillation phenomenon can be observed both in the master equations for the 1st DQD qubit (Fig. \ref{inverse_rabi}) and the corresponding output single-photon distributions (Fig. \ref{dec8}); see the curves for $\Gamma_1=3\times10^{-3}\omega_0$ and $\Gamma_1=5\times10^{-3}\omega_0$ in Figs. \ref{inverse_rabi}-\ref{dec8}. This reveals consistency between the atomic excitation and the output photon distribution.

\begin{figure}[htp!]
\centering
\includegraphics[scale=0.6]{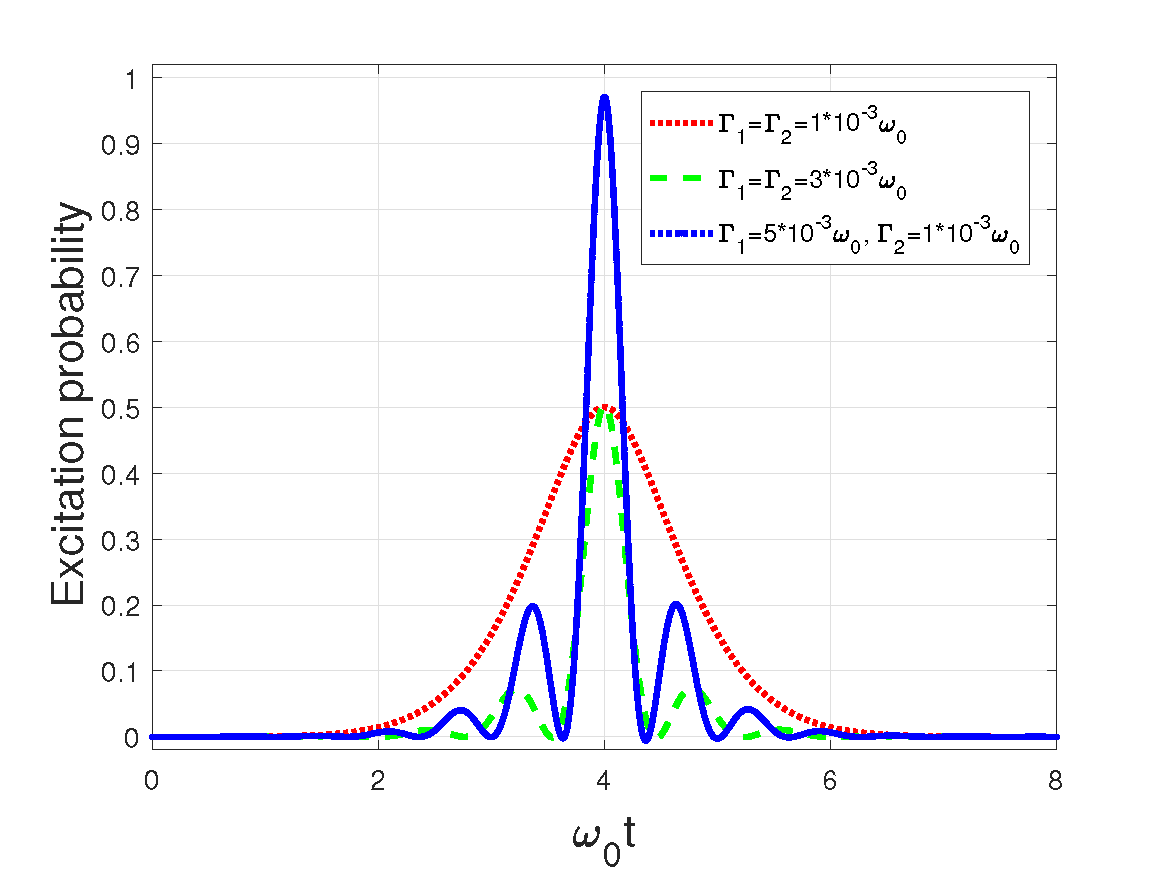}
\caption{The excitation probability of the 1st DQD qubit, we fix the terminal time $t=4$ and choose $\mu=0.2$ and $\kappa=7.5\times10^{-3}\omega_0$.}
\label{inverse_super}
\end{figure}

Finally, let the pulse shape of the single-photon input state be
\begin{equation}\label{Sep26-1}
\ket{\Phi_1}=\left[\gamma_1B^\dagger(\tilde{\nu}_1)+\gamma_2B^\dagger(\tilde{\nu}_2)\right]\ket{\Phi_0},
\end{equation}
where $\tilde{\nu}_1$ and $\tilde{\nu}_2$ are those in \eqref{Sep22-3}, and the complex numbers $\gamma_1$ and $\gamma_2$ satisfy $|\gamma_1\Gamma_1+\gamma_2\Gamma_2|^2=\Gamma_1^2+\Gamma_2^2$ so that the state $\ket{\Phi_1}$ in \eqref{Sep26-1} is normalized.  According to Theorem \ref{Sep22-3},    a single-photon input state $B^\dagger(\tilde{\nu}_1)\ket{\Phi_0}$ is able to fully excite the 1st DQD qubit at time $t$ when the 2nd DQD qubit is decoupled from the cavity,  and likewise,  a single-photon input state $B^\dagger(\tilde{\nu}_2)\ket{\Phi_0}$ is able to fully excite the 2nd DQD qubit at time $t$ when the 1st DQD qubit is decoupled from the cavity.  We are interested in how the DQD qubits are excited by a single-photon state that is a superposition of  $B^\dagger(\tilde{\nu}_1)\ket{\Phi_0}$ and  $B^\dagger(\tilde{\nu}_2)\ket{\Phi_0}$, namely $\ket{\Phi_1}$ in \eqref{Sep26-1}.   Actually, by the proof of Theorem \ref{thm:full inversion}, it can be readily shown that the state $\ket{\Psi(t)}$ of the total system at the terminal time $t$ in \eqref{Sep6-2} is
\begin{equation}
\ket{\Psi(t)} = (\gamma_1|e_1g_2\rangle+\gamma_2|g_1e_2\rangle)\ket{0\Phi_0}.
\end{equation}
In other words, the two DQD qubits form a superposition state and  are decoupled from the rest of the coherent feedback network. The excitation probability of the 1st DQD qubit with the single-photon input state \eqref{Sep26-1} is shown in Fig. \ref{inverse_super}. In Fig. \ref{inverse_super}, it can be observed that the maximum value of the excitation probability is only $0.5$ when the two DQD qubits are equally coupled to the cavity. However, the excitation probability can approximately attains $1$ if $\Gamma_1$ is sufficiently larger than $\Gamma_2$, as predicted by Theorem \ref{thm:full inversion}.

\section{Conclusion}\label{sec:conclusion}
In this paper, we have studied in detail the dynamics of a quantum coherent feedback network of  two distant quantum double dot (DQD) qubits which are directly coupled to a cavity. We have derived an exact form of the steady-state output single-photon state when the network is driven by a continuous-mode single-photon input state. By means of the analytic result, we have analyzed the influence of red, blue, and red+blue detunings on the network dynamics. We have also investigated the excitation probabilities of DQD qubits when the network is driven by a single photon of Gaussian pulse shape. On the other hand, when the input is vacuum while the 1st DQD qubit is in its excited state, the analytic expression of the state of the whole system (field plus the network) is derived, which shows that the output field and the two DQD qubits can be entangled. Finally, we have shown explicitly how to design a temporal pulse shape for a single photon so that it can fully excite a DQD qubit in the network.


\end{document}